\documentclass[pre,twocolumn,floatfix,showpacs]{revtex4}
\usepackage{amsmath,amssymb,graphicx,psfrag}
\usepackage{multirow}
\def\bea{\begin{eqnarray}}
\def\eea{\end{eqnarray}}
\newcommand{\rv}{\vec{r}}

\begin{document}

\title{Phase-field-crystal model for fcc ordering}

\author{Kuo-An Wu\footnote{Present address: Department of Materials Science and 
Engineering, Northwestern University, Evanston IL, 60208, USA}}\author{Ari Adland} \author{Alain  
Karma}

\address{
Department of Physics and Center for Interdisciplinary Research on Complex Systems, Northeastern University, Boston, Massachusetts 02115
}

\begin{abstract}
We develop and analyze a two-mode phase-field-crystal model to describe fcc ordering. The model is formulated 
by coupling two different sets of crystal density waves corresponding to $\langle 111\rangle$ and $\langle 200\rangle$   reciprocal lattice vectors, which are chosen to form triads so as to produce a simple free-energy landscape with
coexistence of crystal and liquid phases. The feasibility of the approach is demonstrated with numerical examples of polycrystalline and (111) twin growth. We use a two-mode amplitude expansion to characterize analytically the free-energy landscape of the model,  identifying parameter ranges where fcc is stable or metastable with respect to bcc. In addition, we derive analytical expressions for the elastic constants for both fcc and bcc. Those expressions show that a non-vanishing amplitude of [200] density waves is essential to obtain mechanically stable fcc crystals with a non-vanishing tetragonal shear modulus $(C_{11}-C_{12})/2$. We determine the model parameters for specific materials by fitting the peak liquid structure factor properties and solid density wave amplitudes following the approach developed for bcc [K.-A. Wu and A. Karma, Phys. Rev. {\bf B} 76, 184107 (2007)]. This procedure yields reasonable predictions of elastic constants for both bcc Fe and fcc Ni using input parameters from molecular dynamics simulations. The application of the model to two-dimensional square lattices is also briefly examined.
\end{abstract}


\maketitle{}

\section{Introduction and summary}
\label{sec:intro}
 
The phase-field-crystal (PFC) method has
emerged as an attractive computational approach to simulate the evolution
of crystalline patterns
\cite{Eldetal02,Eldetal04,Eldetal06,Steetal06,Beretal06,Wu09}.
By resolving the crystal density field, it naturally incorporates 
defects and elastic interactions arising from localized and large scale distortions of this field, respectively. Moreover, this method can in principle be used to simulate microstructural evolution on diffusive time scales  
that are much longer than typical time scales accessible by
molecular dynamics (MD) simulations.

Like classical density function theory (DFT), the PFC method is 
based on representing the free-energy of a material
by a functional of its density 
\cite{RamYus79,HayOxt81,Laietal87,HarOxt84,Sin91,SheOxt96a,SheOxt96b}.
However, classical DFT and the PFC method use different functionals to achieve different
goals. Classical DFT seeks a physically realistic mean-field description of the crystal density field $n(\vec r)$ to reproduce quantitatively as
accurately as possible the properties of a material. 
Since $n(\vec r)$ is sharply peaked around mean atomic positions, this generally requires a very large
number of terms in the traditional expansion of the number density as a sum of density waves  
\begin{equation}
 n(\vec r)=n_0\left(1+\sum_{i} u_ie^{i\vec K_i\cdot \vec r}\right), \label{sumK}
 \end{equation}
where each $\vec K_i$ represents a different reciprocal lattice vector (RLV) in this unrestricted sum. In contrast, by using a considerably simplified density functional, the PFC method essentially restricts this sum to a much smaller set of reciprocal lattice vectors in order to simulate efficiently the evolution of the crystal field on length and time scales as large as possible. Recent studies have shown that, despite this loss of realism, PFC models are able to reproduce quantitatively certain key properties that influence microstructural evolution such as the crystal-melt interfacial free-energy \cite{Wu07,Jaaetal09}, the bulk modulus \cite{Jaaetal09}, and grain-boundary energies \cite{Jaaetal09}, which have been computed for the test case of pure Fe.  

Despite this progress, the PFC method has only been developed for a small set of crystal structures. The original formulation of Elder {\it et al.} \cite{Eldetal02,Eldetal04} uses the same free-energy functional as the Swift-Hohenberg model of pattern formation \cite{SwiHoh77,CroHoh93}   
of the form 
\begin{equation}
{\cal F}=\int d\vec r f,\label{calF}
\end{equation}
with the free-energy density 
\begin{equation}
f = 
\frac{\phi}{2}\left[
a + \lambda({q_0}^2+\nabla^2)^2
\right] \phi
+g\frac{\phi^4}{4},\label{onemode}
\end{equation}
where $\phi$ represents the crystal density field.
This one-mode model
essentially truncates the sum (\ref{sumK}) to one set of RLVs with equal magnitude $|\vec K_i|=q_0$  since higher $K$ modes have much smaller amplitude. As a result, it favors crystal structures for which the principal RLVs can form ``triads'' (i.e. closed triangles), which include hexagonal and body-centered-cubic (bcc) ordering in two and three dimensions, respectively. Aside from favoring those structures, triad interactions are essential for solid-liquid coexistence. This is because in a weakly nonlinear expansion of the bulk free-energy density of the form, $f=c_2u^2+c_3u^3+c_4u^4+\dots$ (with $u_i=u$ for all  principal RLVs), triads contribute a cubic term with a negative coefficient $c_3<0$. Since $c_2$ and $c_4$ are both positive, this cubic term is responsible for the existence of a free-energy barrier between the two minima of $f$ corresponding to liquid ($u=0$) and solid ($u_s>0$).

In this paper, we use a ``two-mode'' phase-field-crystal model to model face-centered-cubic (fcc) structures, which has the free-energy density 
\begin{equation}
f=
\frac{\phi}{2}\left[
a +\lambda (\nabla^2+q_0^2)^2((\nabla^2+q_1^2)^2+r_1)\label{twomode}
\right] 
+g\frac{\phi^4}{4}.
\end{equation}
This model truncates the sum (\ref{sumK}) to two sets of RLVs with magnitude $|\vec K_i|=q_0$ and $|\vec K_i'|=q_1$, respectively, where the first set corresponds in general to the principal RLVs and the second to some other set with larger wavevector magnitude; all other RLVs have much smaller amplitude. This construct provides more flexibility to form triad interactions by combining RLVs from those two sets, and hence to describe other crystal structures. We demonstrate this here for fcc ordering, which is obtained by choosing the sets $\{\vec K_i\}$ and $\{\vec K_i'\}$ to correspond to $\langle 111\rangle$ (principal set) and $\langle 200\rangle$ RLVs, respectively, with $q_1/q_0=\sqrt{4/3}$.
While there is in principle freedom in the choice of the second set $\{\vec K_i'\}$ for a given structure, we have chosen this set such that $q_1>q_0$ is as small as possible, as desired for computational efficiency.

The form  (\ref{twomode}) reduces in the limit $r_1=0$ to the free-energy density introduced by Lifshitz and Petrich \cite{LifPet97} as a generalization of the Swift-Hohenberg model to describe two-dimensional quasiperiodic patterns observed in Faraday wave experiments, which result from the superposition of two frequencies. Although formulated primarily to describe those patterns, this model was also shown to describe other patterns, including regular square crystal lattices in two dimensions with the choice $q_1/q_0=\sqrt{2}$, which couples $\langle 10\rangle$ and $\langle 11\rangle$ RLVs. 

The present introduction of the parameter $r_1$ in the form  (\ref{twomode}) provides the additional flexibility to change the relative stability of different crystal structures. This is because in the limit $r_1\gg q_0^4$, this form reduces formally to the original Swift-Hohenberg form (\ref{onemode}) after a simple rescaling of the parameters. Hence, as $r_1$ is increased the contribution of the second $q_1$-mode becomes less significant in comparison to the first $q_0$-mode. Consequently, as $r_1$ is increased from zero, the crystal structure favored by the two-mode interaction becomes metastable with respect to the one-mode structure. This added capability to model the coexistence of two different crystal structures, in addition to the coexistence of each structure with a liquid, should prove useful to model a wide range of phase transformations with a PFC approach.

In the next section, we scale the parameters of the model to write the free-energy functional in a dimensionless form with only three parameters: $\epsilon$, which is the standard PFC model parameter analogous to temperature that controls the size of the solid-liquid coexistence regions as a function of density, $Q_1\equiv q_1/q_0$, whose value is generally determined by the choice of crystal structure, and $R_1\equiv r_1/q_0^4$ controls the relative stability of the two-mode and one-mode structures (fcc and bcc, respectively). In this section, we also use a standard common tangent construction to compute the phase-diagram in the plane of density and $\epsilon$ for an illustrative choice of $R_1=0.05$. The phase diagram exhibits regions of bcc-liquid and fcc-liquid coexistence for small and large epsilon, respectively. The size of the fcc-liquid coexistence region depends generally on $R_1$. For $r_1=0$ where Eq. (\ref{twomode}) reduces to the free-energy density of Lifshitz and Petrich \cite{LifPet97}, the analog phase-diagram only exhibits fcc-liquid coexistence, so that a finite $r_1$ is necessary for the phase diagram to exhibit both bcc-liquid and fcc-liquid coexistence. We demonstrate the feasibility of the approach with some simulations of polycrystalline growth and (111) twin growth. A numerical computation of the (111) twin boundary energy for parameters of Ni is given in an appendix. The ability to model twin growth is important for solidification modeling since twins can dramatically alter  both eutectic \cite{DayHel68,Napetal2004} and dendritic \cite{Henetal04} microstructures. 

In section III, 
we carry out an amplitude expansion of the bulk free-energy density in the small $\epsilon$ limit. 
This expansion exploits the property that, with the scaling $R_1=\epsilon R$, 
the amplitudes of the $\langle 111\rangle$ and $\langle 200\rangle$ density waves scale as, $ A_s\sim A\epsilon^{1/2}$ and $B_s\sim B\epsilon^{1/2}$, respectively, while the density difference between solid and liquid scales $\sim \epsilon^{3/2}$. Therefore, this density difference can be neglected in the small $\epsilon$ limit and the bulk free-energy density can be
expressed solely in terms of those amplitudes. As required for solid-liquid coexistence, the free-energy density has minima in the ($A$,$B$) plane corresponding to liquid ($A=B=0$) and fcc solid (finite $A$ and $B$ values that depend on $R$). By comparing this form to the free-energy density for a single amplitude of bcc density waves (corresponding to $\langle 110\rangle$ RLVs), we identify different regions of relative fcc and bcc stability, which explains the phase diagram computed in section II.

In section IV, we discuss how to determine the two-mode PFC model parameters to relate them quantitatively to different materials.
We follow essentially the same approach developed by two of the authors for the standard PFC one-mode model for bcc ordering \cite{Wu07}.
For bcc, the parameters were completely determined by fitting three 
parameters: (i) the peak value of the liquid structure factor, $S(q_0)$, where $q_0=|\vec K_{110}|$, (ii) the second derivative of the fourier transform of the direct correlation function at this peak, $C''(q_0)$, and (iii) the solid  
density wave amplitude $u_{110}$. For fcc, all the parameters except $R_1$ are determined by the same fit, where $q_0=|\vec K_{111}|$.  (The shape of the structure factor at $q_1=|\vec K_{200}|$ is not realistically modeled given the limited number of model parameters.) 
$R_1$ then determines the ratio $u_{200}/u_{111}$ of the $\langle 111\rangle$ and $\langle 200\rangle$ solid amplitudes, which can be 
varied to alter the relative stability of fcc and bcc. 

In section V, we derive analytical expressions for the three independent elastic constants of a cubic material, $C_{11}$, $C_{12}$, and $C_{44}$,
for both the standard one-mode PFC model (\ref{onemode}) and the present two-mode model (\ref{twomode}). 
We use a brute force approach that consists of calculating to
quadratic order the change of solid free-energy density, modeled by a one- or two-mode approximation for bcc and fcc, respectively, 
due to small dilation or shear transformations of the unit cell. We have checked that we obtain identical expressions to those derived recently
by Spatschek and Karma for general lattices using an amplitude equation framework \cite{SpaKar09}, which provides a non-trivial self-consistent
test of our calculations. For the one-mode bcc model (\ref{onemode}), the elastic constants are
\begin{equation}
\frac{C_{11}}{2}=C_{12}=C_{44}=-\frac{n_0k_BT}{2}C''(q_0)q_0^2u_{110}^2,\label{C11onemode}
\end{equation}
where $q_0=|\vec K_{110}|$. For the two-mode fcc model (\ref{twomode}),
\begin{equation}
C_{11}=-\frac{4n_0k_BT}{9}C''(q_0)q_0^2\left(u_{111}^2+4u_{200}^2\right), \label{C11twomode}
\end{equation}
and
\begin{equation}
C_{12}=C_{44}=-\frac{4n_0k_BT}{9}C''(q_0)u_{111}^2,\label{C12twomode}
\end{equation}
where $q_0=|\vec K_{111}|$ and $R_1=0$ for simplicity.
 
Using values of $C''(q_0)$ and density wave amplitudes from molecular dynamics 
simulations for parameters of bcc Fe and fcc Ni, 
we find that the above expressions give reasonable estimates of elastic constants
(e.g., $C_{11}\approx 90$ GPa for one-mode bcc PFC model compared to $C_{11}\approx 128$ GPa
in MD Fe and $C_{11}\approx 106$ GPa for the two-mode fcc PFC model compared to $C_{11}\approx 155$ GPa in MD Ni).
The predicted values generally tend to be lower than the constants computed from MD simulations,
but such discrepancies are to be expected given the PFC models are based on one or two modes. 

The analytical predictions for the elastic constants allow us to draw two important general conclusions pertaining to the development of 
PFC models for different crystal structures and to the method used to determine the parameters of those models.

The first conclusion, which follows directly from Eqs. (\ref{C11twomode}) and (\ref{C12twomode}), is that the presence of
the second mode, which corresponds to [200] density waves, is essential to obtain a physically meaningful set of
elastic constants for fcc. Without this second mode ($u_{200}=0$), Eqs. (\ref{C11twomode}) and (\ref{C12twomode}), predict
that $C_{11}=C_{12}=C_{44}$. This implies that the tetragonal shear modulus $C'=(C_{12}-C_{22})/2$ vanishes,
and that the system is mechanically marginally stable. Of course, these analytical expressions for the
elastic constants neglects the contributions of higher modes that are present in a full solution of the PFC equations.
However, those higher modes are generally small for the small values of $\epsilon$ corresponding to
Fe and Ni parameters. Therefore, the contributions of those modes will generally be small and will not change
qualitatively this picture. 

While it is in principle possible to select energetically different crystal structures in the PFC model with
the addition of other nonlinearities in the free-energy density (such as $|\nabla \phi|^4$ and $\phi^2|\nabla \phi|^2$) \cite{Wuthesis}, 
this approach will be of limited applicability for crystal structures like fcc where one mode does not suffice to 
produce the correct elastic properties. This is also true for simple cubic lattices and two-dimensional square lattices. The latter are briefly
examined in section VI by coupling $\langle 10\rangle$ and $\langle 11\rangle$ density waves.

The second conclusion, which is general, is that the elastic constants are
uniquely determined once the phase-field model parameters have been fitted to the peak liquid structure properties, which fixes $C''(q_0)$, 
and the solid density wave amplitudes, as in the approach of Wu and Karma \cite{Wu07} summarized above. 
This also fixes the value of the elastic bulk modulus
\begin{equation}
K=\frac{C_{11}+2C_{12}}{3}\label{Kdef1}
\end{equation}
In general, the bulk modulus can also be defined from the thermodynamic relation
\begin{equation}
K=V\frac{\partial^2 F}{\partial V^2}=n^2\frac{\partial^2 (F/V)}{\partial n^2},\label{Kdef2}
\end{equation}
where $F$ is the total free-energy, $V$ is the volume, and $n=N/V$ is the number density.
The second equality in the last equation can in principle be used to compute the bulk
modulus directly from the PFC solid free-energy curve ($F/V$ versus $n$), without computing the elastic
constants. For a perfect crystal without vacancy, Eqs. (\ref{Kdef1}) and (\ref{Kdef2}) should
in principle predict the same bulk modulus. However, the two definitions can give 
different predictions for the PFC model because the number of atoms per peak of the crystal density field is not 
constrained to unity. While the average number of atoms per peak will also differ from unity
in a real crystal with vacancies, thereby altering the open-system elastic constants \cite{LarCah85}, 
the vacancy concentration is generally very small even at melting.  
How to meaningfully relate the predictions of Eqs. (\ref{Kdef1}) and (\ref{Kdef2}) for the bulk modulus 
is unclear in the PFC approach that, by construct, does not use a realistic description of the crystal density field, and also does not model vacancy formation explicitly.

Despite these limitations of the PFC approach, Jaatinen {\it et al.} \cite{Jaaetal09} have recently proposed a modified one-mode  
PFC model to remedy the fact that, for the standard one-mode PFC model with the free-energy density (\ref{onemode}),
 the bulk modulus predicted by Eq. (\ref{Kdef2}) is several times smaller than
the experimental value for parameters of bcc Fe. Their model yields a value of the bulk modulus computed through Eq. (\ref{Kdef2}) that is
in better agreement with experiment and also gives an improved prediction of the 
density difference between solid and liquid. It gives similar predictions of crystal-melt interfacial free-energies for bcc Fe as
obtained previously by Wu and Karma using the standard one-mode model \cite{Wu07}.

In the light of Eq. (\ref{C11onemode}), 
it is apparent that any one-mode model that fits the correct peak structure factor properties and 
solid density wave amplitudes should predict the same elastic constants. This is consistent with the fact that
Eq. (\ref{C11onemode}) predicts a shear modulus $C_{44}\approx 45$ GPa for the standard
one-mode model of bcc Fe, which is reasonably close to the value $C_{44}\approx 53$ GPa estimated by 
Jaatinen {\it et al.} \cite{Jaaetal09} from numerical 
shearing experiments in their model for similar input parameters. 

Since elastic constants are a major
determinant of grain boundary energies and long-range interactions between crystal defects, 
reproducing those constants, and hence the bulk modulus predicted by
Eq. (\ref{Kdef1}), appears essential for modeling microstructural evolution. Also requiring that 
Eq. (\ref{Kdef2}) predicts the correct bulk modulus using the solid free-energy curve
may appear desirable. However, the motivation for doing so in the context of simple
PFC models is somewhat less clear given the lack of realism of the crystal density field and the fact
that Eqs. (\ref{C11onemode})-(\ref{C12twomode}) predict reasonable values of
the elastic constants. In fact, any one- or two-mode model with the
same peak liquid structure factor properties and density wave amplitudes will predict essentially the same elastic constants associated with the free-energy cost of lattice distortions, 
and also the same interfacial energies as can be inferred from amplitude equations \cite{Wu07}.  
Since those elastic constants and interfacial energies are the quantities that matter most for modeling
microstructural evolution in a PFC context, we have not found it necessary to formulate the two-mode
PFC model in such a way that the bulk modulus is also correctly predicted  from the solid free-energy curve
using Eq. (\ref{Kdef2}). Accordingly, we follow essentially the same approach outlined in Ref. \cite{Wu07} for determining the PFC model parameters.

\section{Phase-field crystal model}

\subsection{Basic equations and scalings}

The PFC equations have the standard form for conserved dynamics  
\bea
\frac{\partial \phi}{\partial \tau} = \Gamma \nabla^2 \frac{\delta {\cal F}}{\delta \phi},
\eea
where ${\cal F}$ is the free-energy functional defined by Eq. (\ref{calF}) with the free-energy densities
given by Eqs. (\ref{onemode}) and (\ref{twomode}) for the one- and two-mode models, respectively.
To minimize the number of parameters,
it is useful to rewrite the equations in dimensionless form. For the two-mode model,
we define the dimensionless parameters
\bea
\label{eq:epsc}
\epsilon =- \frac{a}{\lambda q_0^8},
\eea
\bea
R_1=\frac{r_1}{q_0^4},
\eea
\bea
Q_1=\frac{q_1}{q_0},
\eea
where we set $Q_1 = |\vec K_{200}|/|\vec K_{111}|=\sqrt{4/3}$ for fcc (i.e. $Q_1$ equal to the ratio of the magnitudes of the $\langle 200\rangle$ and $\langle 111\rangle$ RLVs). We also define the
dimensionless variables
\bea
\vec{r'}=q_0\,\vec r,\label{q0sub}
\eea
\bea
\label{eq:ampc}
\psi=\sqrt{\frac{g}{\lambda q_0^8}} \, \phi,
\eea
\begin{equation}
t=\Gamma \lambda q_0^7 \tau,
\end{equation}
\begin{equation}
F=\frac{g}{\lambda^2 q_0^{13}} {\cal F}.
\end{equation}
Substituting the above definitions into Eqs. (\ref{calF}) and
(\ref{twomode})  yields the dimensionless form  
\bea
\frac{\partial \psi}{ \partial t} = \nabla^2 \frac{\delta F}{\delta \psi},\label{dyn}
\eea
with the free-energy functional  
\begin{equation}
F=\int d \vec{r}f(\psi),\label{Fdef}
\end{equation}
and free-energy density
\begin{equation}
f=  \frac{\psi}{2}\left[ -\epsilon +(\nabla^2+1)^2
((\nabla^2+{Q_1}^2)^2+R_1) \right] \psi 
 +\frac{\psi^4}{4},\label{twomode2}
\end{equation}
where we have dropped the prime symbol on the dimensionless spatial
coordinate vector $\vec{r'}$ for brevity. Even though most of the paper focuses on the
two-mode model, we also compute in section V the elastic constants for the standard one-mode PFC model.
For this model, we use the same scaling as
in Ref. \cite{Wu07} with the parameter
\bea
\label{epsonemode}
\epsilon =- \frac{a}{\lambda q_0^4},
\eea
and dimensionless variables 
\bea
\psi=\sqrt{\frac{g}{\lambda q_0^4}} \, \phi,
\eea
\begin{equation}
t=\Gamma \lambda q_0^3 \tau,
\end{equation}
\begin{equation}
F=\frac{g}{\lambda^2 q_0^5} {\cal F},
\end{equation}
where $\vec{r'}$ is defined by Eq. (\ref{q0sub}).
Substituting the above forms into Eqs. (\ref{calF}) and
(\ref{onemode})  yields (after dropping the prime symbol on $\vec{r'}$)  
the dimensionless form of the one-mode PFC 
equations (\ref{dyn}) and (\ref{Fdef}) with 
\bea
f=  \frac{\psi}{2}\left[ -\epsilon +(\nabla^2+1)^2\right] \psi 
 +\frac{\psi^4}{4}.\label{onemode2} 
\eea

\subsection{Phase diagram}

The phase diagram of the two-mode PFC model is obtained  
by computing the free-energy density 
as a function of the mean density
$\bar \psi$ in solid and liquid, denoted by $f_s(\bar \psi)$, 
and $f_l(\bar \psi)$, respectively, and then using a standard
common tangent construction to obtain
equilibrium values of $\bar \psi$ in 
solid ($\bar \psi_s$) and liquid ($\bar \psi_l$).

\begin{figure}
\centering
\includegraphics[width=0.5\textwidth, angle=0]{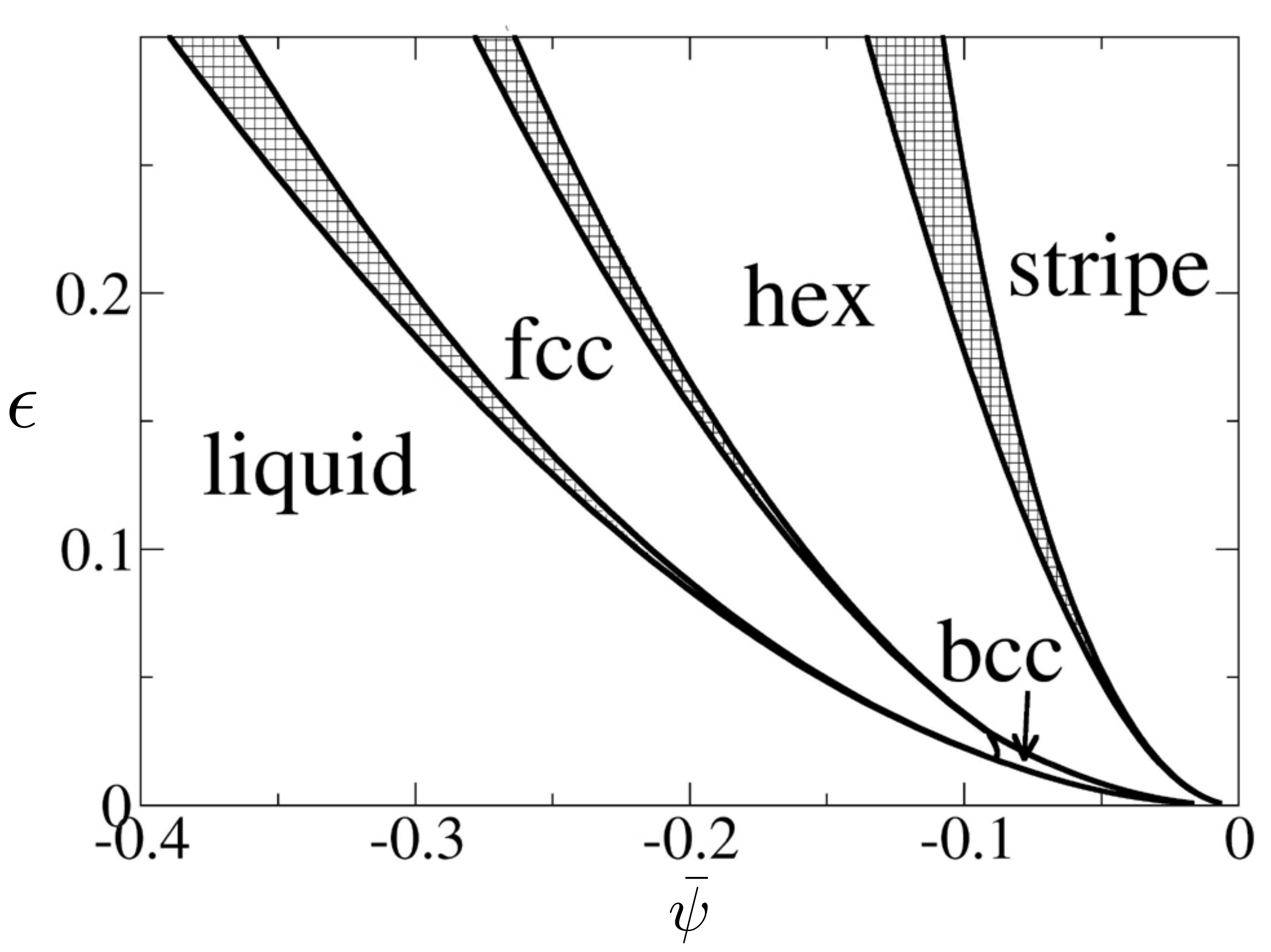}
\caption{Phase diagram of the two-mode PFC model for $R_1 = 0.05$ computed using two-mode and one-mode expansions of the crystal density field for fcc and bcc, respectively.}
\label{phase}
\end{figure}

\begin{figure}
\centering
\includegraphics[width=0.5\textwidth, angle=0]{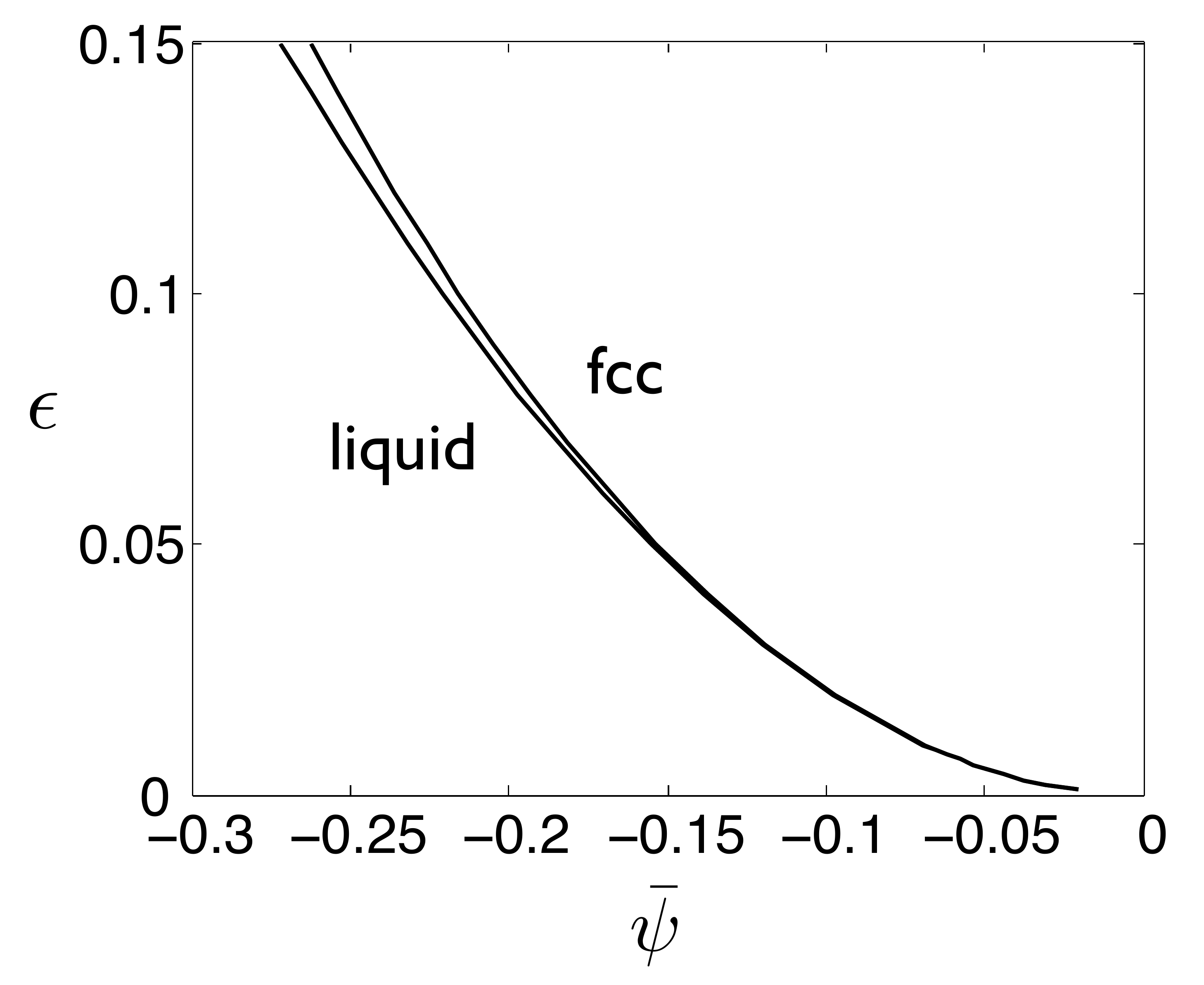}
\caption{Phase diagram of the two-mode PFC model showing only the fcc-solidus and liquidus for the case $R_1 = 0$ where fcc-liquid coexistence extends to vanishingly small $\epsilon$.}
\label{phaser0}
\end{figure}

Since the density is constant in the liquid, $f_l$ 
is obtained directly from
Eq. (\ref{twomode2})    
\begin{equation}
f_l(\bar{\psi}_l)=-(\epsilon-\frac{16}{9}-R_1)\frac{\bar\psi_l^2}{2}+\frac{\bar\psi_l^4}{4}.
\label{eq:fldef}
\end{equation}
For small $\epsilon$,  
the solid free-energy density 
can be well approximated by only considering the contribution
of the $\langle 111\rangle$ and $\langle 200\rangle$ RLVs. Accordingly, 
the crystal density field is expanded in the form
\begin{eqnarray}
\label{eq:fccs}
\psi(\vec{r})&\approx& \bar\psi  + \sum_{\vec{K_i}=\langle 111\rangle} A_i\, e^{i\vec{K_i}\cdot \vec{r}} 
 + \sum_{\vec{K_j}'=\langle 200\rangle} B_j\, e^{i\vec{K_j}'\cdot \vec{r}} \nonumber \\
&\approx& \bar\psi + 8A_s\cos{qx}\cos{qy}\cos{qz} \nonumber \\
& & + 
2B_s (\cos{2qx}+\cos{2qy}+\cos{2qz}),
\end{eqnarray}
where we have used the fact that all density waves have the same amplitude in the crystal
($|A_i|=A_s$ and $|B_i|=B_s$) and the magnitude of the principal RLVs are unity in our dimensionless units
so ($q=1/\sqrt{3}$).
The parameters $A_s$ and $B_s$ are solved by substituting Eq.~(\ref{eq:fccs})
 into  Eqs. (\ref{Fdef}) and (\ref{twomode2})
 and by minimizing the resulting free-energy $F$ with respect to
$A_s$ and $B_s$.
This minimization yields the solid free energy density  
\begin{eqnarray}
f_s(\bar\psi_s) &=& 4\,(-\epsilon + 3 {\bar{\psi}_s}^2 ) A_s^2 
+3\, (-\epsilon +  3 {\bar{\psi}_s}^2 + \frac{R_1}{9}  )  B_s^2 \nonumber \\ 
&&+72 \bar{\psi}_s A_s^2 B_s  
+144A_s^2 B_s^2   +54 A_s^4 + \frac{45}{2} B_s^4 \nonumber \\
&&-\frac{\epsilon}{2} {\bar{\psi}_s}^2 + \frac{R_1}{2} {\bar{\psi}_s^2}
+\frac{8}{9}{\bar{\psi}_s}^2 + \frac{1}{4} {\bar{\psi}_s}^4,
\label{eq:fs}
\end{eqnarray}
where $A_s$ and $B_s$ are themselves functions of $\bar{\psi}$.
The coexistence densities $\bar{\psi}_s$ and $\bar{\psi}_l$ are computed numerically using the
standard common tangent construction,
which consists of equating the chemical potentials  
$f'_s(\bar{\psi}_s) = f'_l(\bar{\psi}_l)=\mu_E$ and grand 
potentials $f_s(\bar{\psi}_s)-\mu_E \bar{\psi}_s = f_l(\bar{\psi}_l)-\mu_E \bar{\psi}_l$ of the two phases.   
It is also necessary to compute the solid free-energy curve for bcc since the latter can have a lower free-energy than fcc for some regions of the phase diagram. The bcc free-energy density was obtained by expanding
the crystal density field using a one-mode
approximation, which only involves $\langle 110 \rangle$ RLVs as in Ref. \cite{Wu07}, and substituting this
expansion into the two-mode model defined by  Eqs. (\ref{Fdef}) and (\ref{twomode2}).

\begin{figure}
\centering
\includegraphics[width=0.4\textwidth, angle=0]{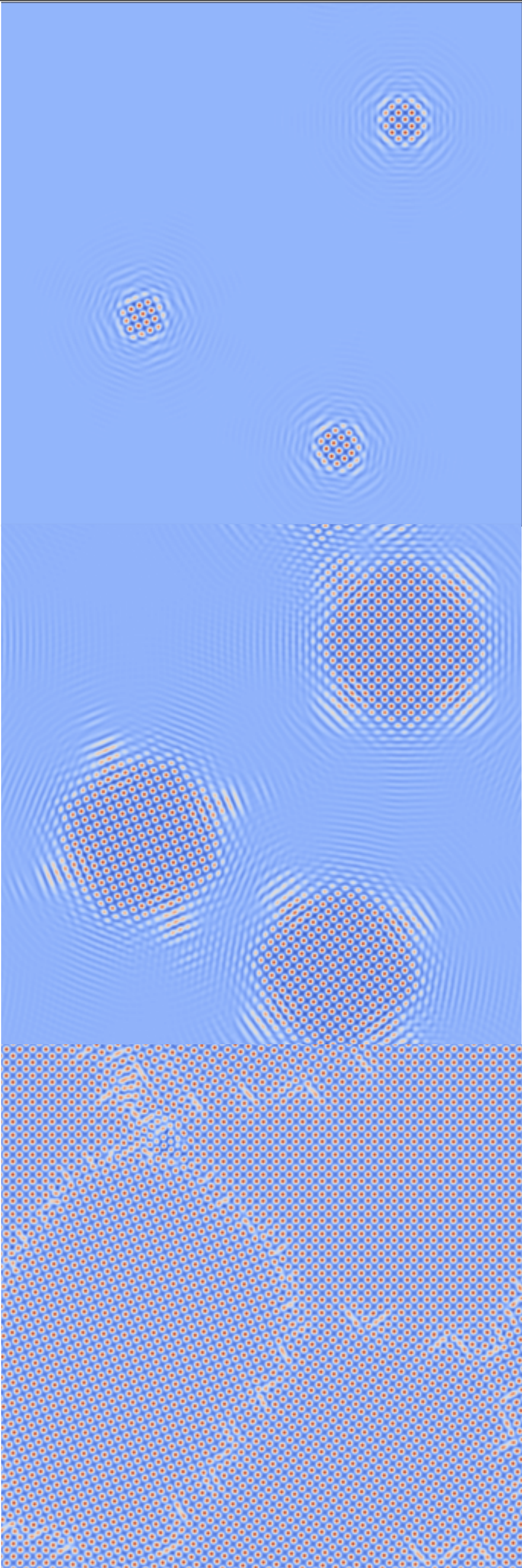}
\caption{Simulation of polycrystalline solidification starting from three seeded fcc crystals in a supercooled liquid.  The system is fully periodic, and the snapshots are taken at dimensionless times $t=5\times 10^2, 3\times 10^3,\mbox{and}\,10^5$.  The parameters are $R=0$, $\epsilon=.00823$, and $\bar{\psi}=-0.06$.}
\label{poly}
\end{figure}

An example of the phase diagram for $R_1=0.05$ is shown in Fig.~\ref{phase}, where we also show for completeness the 
hexagonal and stripe phases. As desired, we obtain a large $\epsilon$ range of fcc-liquid coexistence. For small $\epsilon$, however, bcc becomes favored over fcc. A common tangent construction using fcc and bcc free-energy curves shows that the density range of bcc-fcc coexistence is extremely narrow for small values of $\epsilon$ and cannot be resolved on the scale of Fig. \ref{phase}. As will be explained later in section \ref{stabfccbcc}, the range of $\epsilon$ where bcc is favored depends on the value of $R_1$. In the limit $R_1\gg 1$, the two-mode model reduces to the standard one-mode model after a simple rescaling of parameters, which can be easily seen by comparing Eqs. (\ref{twomode2}) and
(\ref{onemode2}). Hence, increasing $R_1$ reduces the contribution of the second mode. Conversely, reducing $R_1$ increases the contribution of this mode and tends to favor the fcc structure, which extends to smaller $\epsilon$ for smaller $R_1$. In the extreme case where $R_1=0$, the region of fcc-liquid coexistence extends all the way to vanishingly small $\epsilon$ as shown in Fig. \ref{phaser0}.

\subsection{Numerical examples}

We now demonstrate the feasibility of the model  
with some numerical examples of fcc polycrystalline growth and (111) twin growth. The PFC conserved dynamics governed by Eq. (\ref{dyn}) with the free-energy defined by Eqs. (\ref{Fdef}) and (\ref{twomode2}) was solved using the semi-implicit pseudo-spectral scheme given by Eq. (A2) in Appendix A of Ref. \cite{Mathis}. We used the parameters $R=0$ and 
$\epsilon=0.00823$ obtained from our fit of pure Ni presented later in section IV, together with the grid spacing $\Delta x =\Delta y =\Delta z = 2\pi \sqrt{3}/16$, which determines the number of Fourier modes, and
the time step $\Delta t=0.5$. For this value of $R$ and $\epsilon$, the computations presented in the next section show that the size of the solid-liquid coexistence region is extremely small, i.e. $\bar\psi_s-\bar\psi_l$ is two orders of magnitude smaller than $(\bar \psi_s+\bar \psi_l)/2$ as can already be seen from the phase diagram in Fig. \ref{phaser0}, and $\bar\psi_s\approx \bar \psi_l \approx -0.0627$.

\begin{figure}
\includegraphics[width=.5\textwidth, angle=0]{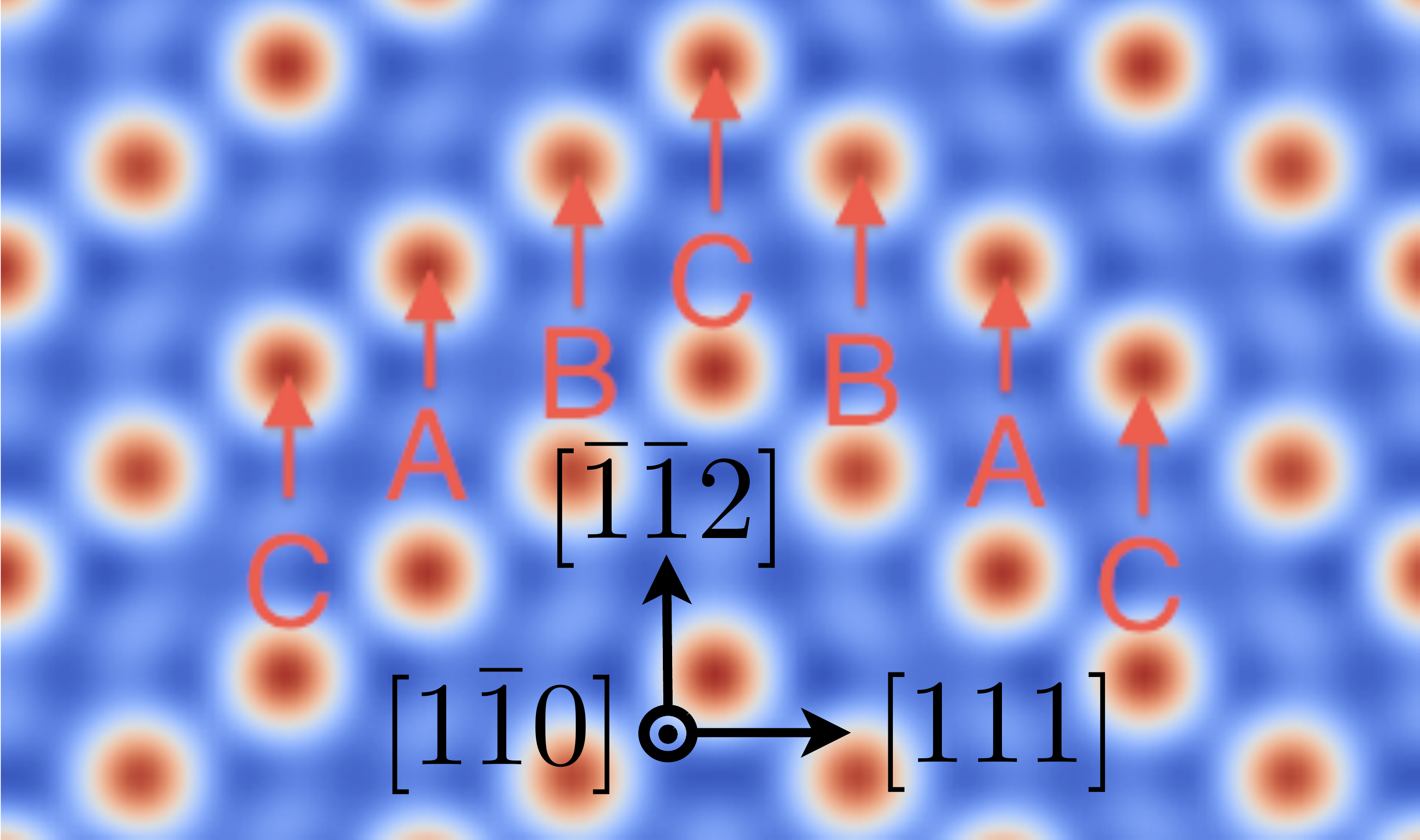}
\caption{Simulation of an equilibrium coherent (111) twin boundary for $R=0$, $\epsilon=0.00823$, and $\bar{\psi}=-0.06269$.}
\label{twin}
\end{figure}

\begin{figure}
\includegraphics[width=.5\textwidth, angle=0]{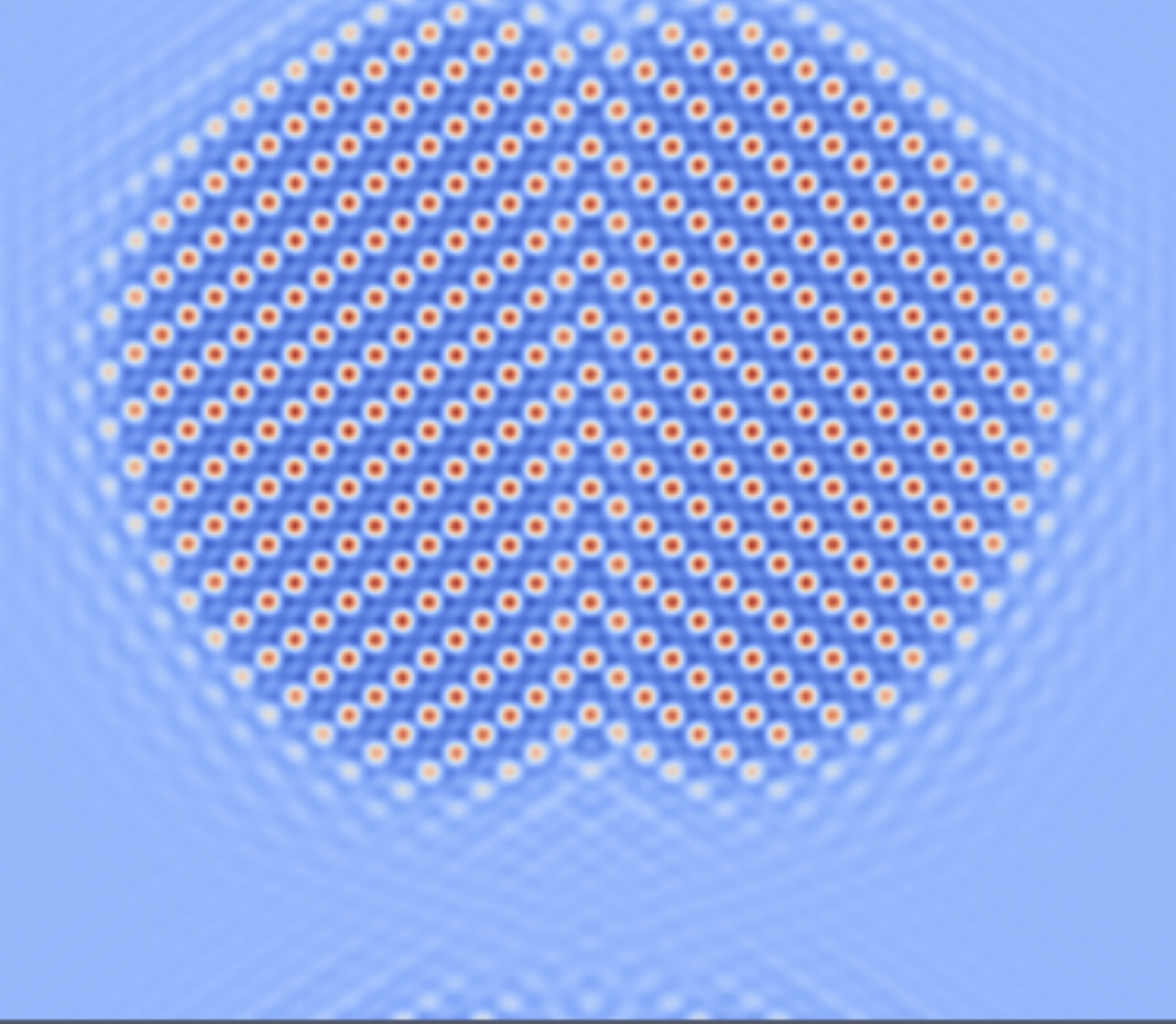}
\caption{Simulation of the growth of a twin crystal in a supercooled liquid for $R=0$, $\epsilon=0.00823$, and $\bar{\psi}=-0.06$.}
\label{twingrow}
\end{figure}

The first example in Fig. \ref{poly} shows the growth of small fcc crystallites of different orientations for a value of $\bar \psi=-0.06>\bar \psi_s$ that is well inside the stable fcc-solid region of the phase diagram. The crystallites grow as expected until they collide to form grain boundaries. The second example in Fig. \ref{twin} shows a (111) twin crystal for a value of
$\bar \psi=-0.06269$ at coexistence and for a system size chosen such that a twin crystal with two stacking faults fits perfectly the periodic boundary conditions in all directions without any liquid present. A computation of the excess free-energy of this twin boundary given in the appendix to this paper yields a value of approximately 30 mJ/m$^2$ that falls within the range of values typically reported in the literature for fcc metals. Fig. \ref{twingrow} then shows the growth of the same twin crystal in a supercooled liquid for a much larger system with $\bar \psi=-0.06$.

\section{Amplitude equations}

\subsection{Scalings}

In this section, we analyze in more detail the properties of the model by expanding the free-energy in terms of the amplitudes of density waves. In the one-mode bcc case analyzed in Ref. \cite{Wu07}, a similar expansion exploited the fact that the amplitude of $\langle 110 \rangle$ density waves scales as $\epsilon^{1/2}$ in the small $\epsilon$ limit. In the present case, the expansion is rendered more difficult by the presence of two different sets of  density waves with amplitudes $A_s$ and $B_s$ corresponding to $\langle 111 \rangle$ and $\langle 200 \rangle$ RLVs, respectively. Therefore, it is not {\it a priori} obvious how $A_s$ and $B_s$ should scale in the small $\epsilon$ limit. If $R_1$ is kept constant, the bcc structure turns out to always be favored in the small $\epsilon$ limit as apparent in the phase diagram of Fig. \ref{phase}. Consequently, a small $\epsilon$ amplitude expansion that captures the fcc structure cannot be carried out at fixed $R_1$. However, if $R_1$ is decreased proportionally to $\epsilon$ by imposing the additional scaling $R_1=\epsilon R$, both $A_s$ and $B_s$ scale as  
$\epsilon^{1/2}$, thereby making a rigorous expansion possible. This expansion may seem artificial since the phase diagram of Fig. \ref{phase} is computed at fixed $R_1$. However, as we show below, the results of this expansion can be used to understand the small $\epsilon$ structure of the phase diagram, in particular the relative stability of fcc and bcc. 

\begin{figure}
\includegraphics[width=.5\textwidth, angle=0]{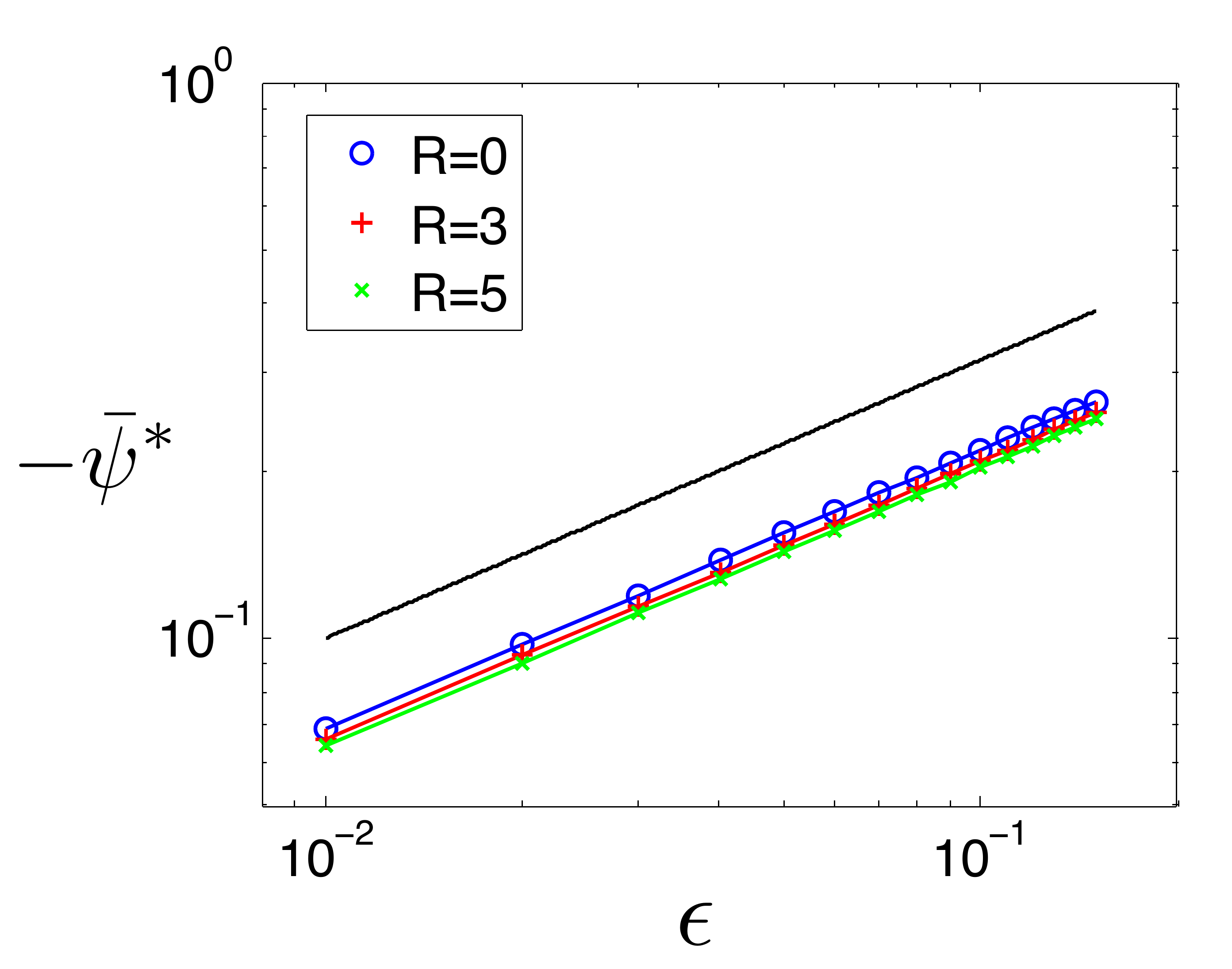}
\caption{(Color Online) Plots of $\bar{\psi}^* \equiv \frac{1}{2}
(\bar{\psi}_s+\bar{\psi}_l)$ versus $\epsilon$ for different values of R, with $\bar{\psi}_s$ and $\bar{\psi}_l$ calculated from the common tangent construction.  Fits to the numerical results of the form $\bar{\psi}^*=\psi_c\epsilon^{1/2}$ yield for $R=0$, $\psi_c=-0.6901$, for $R=3$, $\psi_c=-0.6578$, and for $R=5$, $\psi_c=-0.6396$. The solid black line has a slope of exactly $1/2$ on this log-log plot showing that the mean density $\sim \epsilon^{1/2}$.}
\label{fcc_liq_avg_psi}
\end{figure}

\begin{figure}
\centering
\includegraphics[width=0.5\textwidth, angle=0]{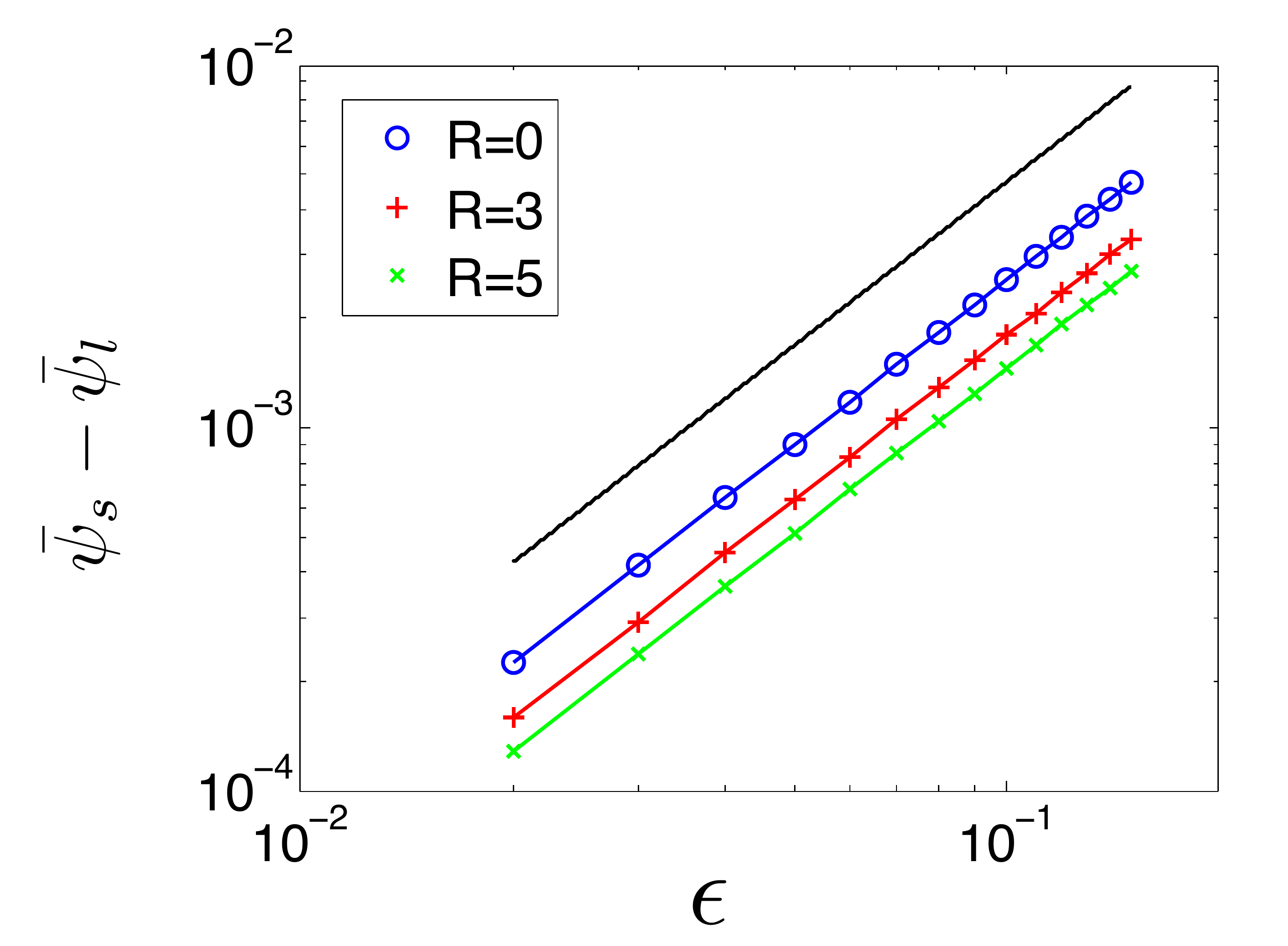}
\caption{(Color Online) Plots of $\bar{\psi}_s-\bar{\psi}_l$ versus $\epsilon$ for the same values of R as in Fig. \protect\ref{fcc_liq_avg_psi}, with  $\bar{\psi}_s$ and $\bar{\psi}_l$ calculated from the common tangent construction. The solid black line has a slope of exactly $3/2$ on this log-log plot showing that the density difference between solid and liquid $\sim \epsilon^{3/2}$.}
\label{loglog}
\end{figure}

To demonstrate the feasibility of this expansion, we first analyze 
fcc-liquid coexistence for small $\epsilon$ with the scaling
$R_1 = \epsilon R$.
The equilibrium densities are calculated using the common tangent
construction described in the previous section.
To make the dependence of the coexistence densities on $\epsilon$ explicit, we make a log-log plot of the mean coexistence density $\bar{\psi}^*\equiv \frac{1}{2} (\bar{\psi}_l+\bar{\psi}_s)$
versus $\epsilon$ for three different values of $R$.
The results in Fig.~\ref{fcc_liq_avg_psi} show that the mean coexistence density scales as $\epsilon^{1/2}$. Next in Fig.~\ref{loglog}, we show a log-log plot of the density difference between solid and liquid versus $\epsilon$ for the same three values of $R$.  The results show that $\bar \psi_s-\bar \psi_l\sim \epsilon^{3/2}$. 
Together, these two log-log plots show that, in the small $\epsilon$ limit, the two-mode PFC model
exhibits a weak first-order freezing transition where the size of the
solid-liquid coexistence region is at the order of $\epsilon^{3/2}$ that is
much smaller than the mean value of the density $\sim \epsilon^{1/2}$.

\subsection{Free-energy functional}

\begin{figure}
\centering
\includegraphics[width=0.5\textwidth, angle=0]{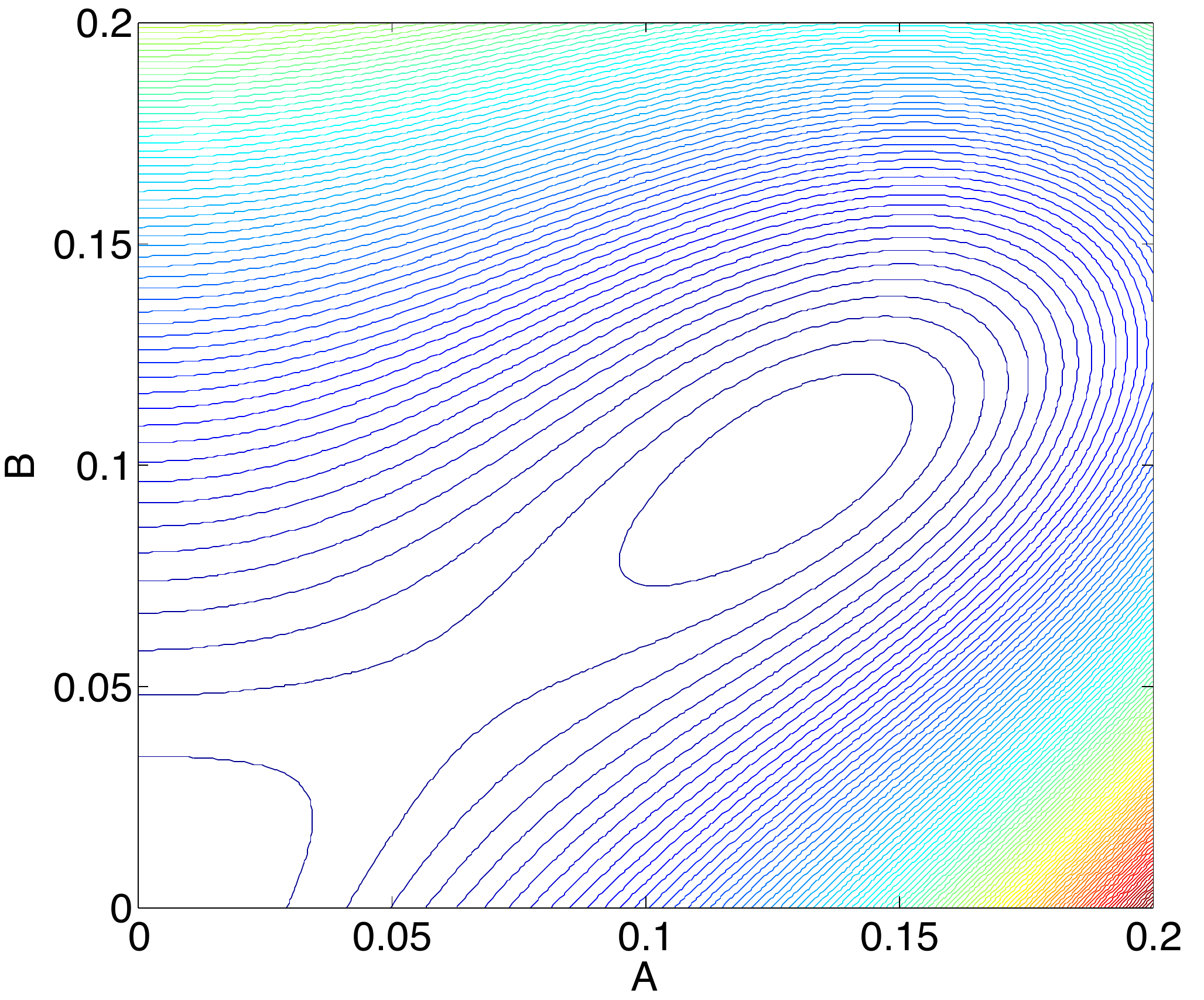}
\caption{(Color online) Free-energy landscape defined by Eq. (\ref{eq:feneq}) as a function of the amplitudes $A$ and $B$ of $\langle 111 \rangle$ and  $\langle 200 \rangle$ density waves, respectively, for $R=0$ and the corresponding coexistence value $\psi_c=-0.6901$ where the solid and liquid minima have the same height.}
\label{F_AB}
\end{figure}

The above scalings suggest that we can expand the crystal density field
in powers of $\epsilon^{1/2}$ as
\bea
\label{eq:den_exp}
{\psi(\vec{r})} &=& {\psi_0(\vec{r})} \epsilon^{1/2} + {\psi_1(\vec{r})} \epsilon + 
\psi_2(\vec{r}) \epsilon^{3/2} + \cdots, 
\eea
and expand accordingly the average densities
\begin{eqnarray}
\bar{\psi}_l &=& {\psi_{l0}} \epsilon^{1/2} + {\psi_{l1}} \epsilon + 
{\psi_{l2}} \epsilon^{3/2} + \cdots, 
\end{eqnarray}
and
\begin{eqnarray}
\bar{\psi}_s = {\psi_{s0}} \epsilon^{1/2} + {\psi_{s1}} \epsilon + 
{\psi_{s2}} \epsilon^{3/2} + \cdots,
\end{eqnarray}
in the liquid and solid, respectively.
The numerically determined scaling relations $(\bar{\psi}_l+\bar{\psi}_s)/2\sim \epsilon^{1/2}$ and
$\bar{\psi}_s-\bar{\psi}_l\sim \epsilon^{3/2}$ then imply that
\begin{eqnarray}
{\psi_{l0}}={\psi_{s0}}\equiv \psi_c,
\end{eqnarray}
and
\begin{eqnarray}
{\psi_{l1}}={\psi_{s1}}=0.
\end{eqnarray}
Next, to carry out the amplitude expansion, we start from the equilibrium
equation $\delta F/\delta \psi=\mu_E$, where $\mu_E$ is the equilibrium value of
the chemical potential. With $F$ defined by Eqs. (\ref{Fdef}) and (\ref{twomode2}), we obtain
\begin{eqnarray}
\label{mu}
\mu_E = -\epsilon \psi +
(\nabla^2+1)\left[(\nabla^2+Q_1^2)^2 +\epsilon R \right] \psi + \psi^3.\,\,\,
\end{eqnarray}
We substitute the small $\epsilon$ expansion of the density field  (\ref{eq:den_exp}) 
into  Eq. (\ref{mu}) and collect terms with the same power of $\epsilon$.
We find at the order $\epsilon^{1/2}$
\begin{eqnarray}
(\nabla^2+1)^2(\nabla^2+Q_1^2)^2 \psi_0 = Q_1^4 \psi_c,
\end{eqnarray}
which has the solution
\begin{eqnarray}
\psi_0 = \psi_c + \sum_i A_i^0 e^{i \vec{K_i}\cdot \vec{r}} + 
\sum_j B_j^0 e^{i \vec{K_j}'\cdot \vec{r}},
\end{eqnarray}
where the summations are over $\langle 111\rangle$ and $\langle 200 \rangle$ RLVs, respectively,
and $|\vec{K_i}|=1,|\vec{K_j}'|=\sqrt{4/3}$, in our scaled units.
At order $\epsilon$, we obtain
\begin{eqnarray}
(\nabla^2+1)^2(\nabla^2+Q_1^2)^2 \psi_1 = 0,
\end{eqnarray}
which has the solution
\begin{eqnarray}
\psi_1 = \sum_i A_i^1 e^{i \vec{K_i}\cdot \vec{r}} + 
\sum_j B_j^1 e^{i \vec{K_j}'\cdot \vec{r}},
\end{eqnarray}
and collecting the terms at order $\epsilon^{3/2}$ yields
\begin{eqnarray}
\label{thorder}
& &-\psi_0+ (\nabla^2+1)^2(\nabla^2+Q_1^2)^2 \psi_2 + R(\nabla^2+1)^2\psi_0 + 
(\psi_0)^3 \nonumber \\
& & 
=-\psi_c + Q_1^4 \psi_{l2} +     R \psi_c +(\psi_c)^3.
\end{eqnarray}
Since $(\nabla^2+1)^2(\nabla^2+Q_1^2)^2 \psi_2$ gives a vanishing contribution
for all density waves associated with sets $\{K_i\}$ and $\{K_j'\}$, all remaining
terms $\sim e^{i \vec{K_i}\cdot \vec{r}}$ and $e^{i \vec{K_j}'\cdot \vec{r}}$ must
balance each other in order for a solution of Eq.~(\ref{thorder})
 to exist.
For example, the condition that the coefficients of $e^{i \vec{K_{111}} \cdot \vec{r}}$
balance each other yields
\bea
\label{equiv_A}
&&(-1+ 3{\psi_c}^2 ) A^0_{111} +
(3|A^0_{111}|^2 + 6 |A^0_{1 {\bar{1}} 1}|^2
+ 6 |A^0_{1 1 {\bar{1}}} |^2
\nonumber \\
&&
+ 6 |A^0_{1 {\bar{1}} {\bar{1}}} |^2 
+ 6|B^0_{200}|^2 
+ 6 |B^0_{020}|^2
+ 6 |B^0_{002} |^2
) A^0_{111}
\nonumber \\
&&
+ 6{\psi_c} (A^0_{\bar{1} 11} B^0_{200} +
A^0_{1 \bar{1} 1} B^0_{0 2 0} +
A^0_{1 1 \bar{1} } B^0_{0 0 2 } 
)   
\nonumber \\
&&
+ 6 A^0_{11\bar{1}} A^0_{1\bar{1}1} A^0_{\bar{1}11}
+ 6A^0_{\bar{1}\bar{1}1} B^0_{200} B^0_{020}
\nonumber \\
&&
+ 6A^0_{\bar{1}1\bar{1}} B^0_{200} B^0_{002}
+ 6A^0_{1\bar{1}\bar{1}} B^0_{002} B^0_{020}
=0,
\eea
and requiring that the coefficients of $e^{i \vec{K_{200}} \cdot \vec{r}}$ balance each other yields in turn
\bea
\label{equiv_B}
&&(-1+ 3{\psi_c}^2+R(-Q_1^2+1)^2 ) B^0_{200} +
(6|A^0_{111}|^2 
+ 6 |A^0_{1 {\bar{1}} 1}|^2
\nonumber \\
&&
+ 6 |A^0_{1 1 {\bar{1}}} |^2
+ 6 |A^0_{1 {\bar{1}} {\bar{1}}} |^2 
+ 3|B^0_{200}|^2 + 6 |B^0_{020}|^2
\nonumber \\
&&
+ 6 |B^0_{002} |^2
) B^0_{200}
+ 6 {\psi_c} ( A^0_{11 \bar{1}} A^0_{1 \bar{1} 1}
+ A^0_{111} A^0_{1 \bar{1} \bar{1} })
\nonumber \\
&&
+6 ( B^0_{020}A^0_{1\bar{1}1}A^0_{1\bar{1}\bar{1}}+
B^0_{002}A^0_{1\bar{1}\bar{1}}A^0_{11\bar{1}}
\nonumber \\
&& +
B^0_{0\bar{2}0}A^0_{11\bar{1}}A^0_{111}+
B^0_{00\bar{2}}A^0_{1\bar{1}1}A^0_{111})
 =0.
\eea
The above solvability condition must be satisfied independently for each 
reciprocal lattice vector.
This yields a set of fourteen coupled amplitude equations that are
straightforward to obtain. From those amplitude equations, 
it is useful to express the free-energy of the system measured from its constant value in the
liquid, defined as the difference $\Delta F^{AE}$,
as a functional of the density wave amplitudes $A_i^0$ and $B^0_i$.  
This quantity can be expressed solely in terms of the amplitudes of density waves owing to
the property that the size of the coexistence region ($\sim \epsilon^{3/2}$) is much smaller than
the mean density ($\sim \epsilon^{1/2}$) in the small $\epsilon$ limit.
Since the amplitudes are not conserved order parameters, the equilibrium state
simply corresponds to a minimum of this free-energy. Hence
the amplitude equations must be equivalent to
\begin{eqnarray}
\frac{\delta \Delta F^{AE}}{\delta A^{0*}_i}=0,
\end{eqnarray}
and
\begin{eqnarray}
\frac{\delta \Delta F^{AE}}{ \delta B^{0*}_i}=0.
\end{eqnarray}
For the case where the amplitudes are spatially uniform, we obtain the free-energy density  
\bea
\Delta F^{AE}/V &\equiv & \Delta f^{AE}\label{eq:faeee}\\
&=&
(-1+ 3{\psi_c}^2)  (
|A^0_{111}|^2 + |A^0_{11\bar{1}}|^2 +|A^0_{1\bar{1}1}|^2 
\nonumber \\
&&
+|A^0_{1\bar{1}\bar{1}}|^2 
) 
+(-1+ 3{\psi_c}^2+R(-Q_1^2+1)^2 ) 
\nonumber \\
&&
(
|B^0_{200}|^2 + |B^0_{020}|^2 +|B^0_{002}|^2 
) 
\nonumber \\
&&
+\frac{3}{2} (
|A^0_{111}|^4 
+ |A^0_{11\bar{1}}|^4 
+|A^0_{1\bar{1}1}|^4 +|A^0_{1\bar{1}\bar{1}}|^4
\nonumber \\
&&
+|B^0_{200}|^4 
+ |B^0_{020}|^4 +|B^0_{002}|^4
) 
\nonumber \\
&&
+ 6(
|A^0_{111}|^2 |A^0_{1\bar{1}1}|^2 
+|A^0_{111}|^2 |A^0_{11\bar{1}}|^2
\nonumber \\
&&
 + |A^0_{111}|^2 |A^0_{1\bar{1}\bar{1}}|^2
+|A^0_{1\bar{1}1}|^2 |A^0_{11\bar{1}}|^2 
\nonumber \\
&&
+ |A^0_{1\bar{1}1}|^2 |A^0_{1\bar{1}\bar{1}}|^2
+|A^0_{11\bar{1}}|^2 |A^0_{1\bar{1}\bar{1}}|^2)
\nonumber \\
&&
+ 6(
|B^0_{200}|^2 |B^0_{020}|^2 
+ |B^0_{200}|^2 |B^0_{002}|^2
\nonumber \\
&&
+ |B^0_{020}|^2 |B^0_{002}|^2
) 
+ 6 (
|A^0_{111}|^2 
+ |A^0_{11\bar{1}}|^2 +|A^0_{1\bar{1}1}|^2
\nonumber \\
&&
 +|A^0_{1\bar{1}\bar{1}}|^2
)
(
|B^0_{200}|^2 
+ |B^0_{020}|^2 +|B^0_{002}|^2
) 
\nonumber \\
&&
+ 6 {\psi_c} (
A^0_{111}A^0_{1\bar{1}\bar{1}} B^0_{\bar{2}00}
+A^0_{111}A^0_{\bar{1}1\bar{1}} B^0_{0\bar{2}0}
\nonumber \\
&&
+A^0_{111}A^0_{\bar{1}\bar{1}1} B^0_{00\bar{2}} +A^0_{11\bar{1}}A^0_{1\bar{1}1} B^0_{\bar{2}00} 
\nonumber \\
&&
+ A^0_{11\bar{1}}A^0_{\bar{1}11} B^0_{0\bar{2}0} + A^0_{11\bar{1}}A^0_{\bar{1}\bar{1}\bar{1}} B^0_{002}
\nonumber \\
&&
+A^0_{1\bar{1}1}A^0_{\bar{1} 1 1} B^0_{00\bar{2}} 
+ A^0_{1\bar{1}1}A^0_{\bar{1}\bar{1}\bar{1}} B^0_{020}
\nonumber \\
&&
+ A^0_{1\bar{1}\bar{1}}A^0_{\bar{1}1\bar{1}} B^0_{002} + A^0_{1\bar{1}\bar{1}}A^0_{\bar{1}\bar{1}1} B^0_{020}
\nonumber \\
&&
+A^0_{\bar{1}11}A^0_{\bar{1}\bar{1}\bar{1}} B^0_{200} + A^0_{\bar{1} 1\bar{1}}A^0_{\bar{1}\bar{1} 1} B^0_{200}
) 
\nonumber \\
&& 
+ 6 A^0_{11\bar{1}} A^0_{1\bar{1}1} A^0_{\bar{1}11} A^0_{\bar{1}\bar{1}\bar{1}} 
+ 6 A^0_{\bar{1}\bar{1}1} A^0_{\bar{1}1\bar{1}} A^0_{1\bar{1}\bar{1}} A^0_{111}
\nonumber \\
&&
+ 6A^0_{\bar{1}\bar{1}1} A^0_{\bar{1}\bar{1}\bar{1}} B^0_{200} B^0_{020}
+ 6A^0_{\bar{1}1\bar{1}} A^0_{\bar{1}\bar{1}\bar{1}} B^0_{200} B^0_{002}  \nonumber \\
&&
+ 6A^0_{1\bar{1}\bar{1}} A^0_{\bar{1}\bar{1}\bar{1}} B^0_{002} B^0_{020}
+ 6A^0_{\bar{1}11} A^0_{111} B^0_{00\bar{2}} B^0_{0\bar{2}0}  \nonumber \\
&&
+ 6A^0_{11\bar{1}} A^0_{111} B^0_{\bar{2}00} B^0_{0\bar{2}0}
+ 6A^0_{1\bar{1}1} A^0_{111} B^0_{\bar{2}00} B^0_{00\bar{2}}  \nonumber \\
&&
+ 6A^0_{\bar{1}11} A^0_{\bar{1}1\bar{1}} B^0_{200} B^0_{0\bar{2}0}
+ 6A^0_{1\bar{1}\bar{1}} A^0_{1\bar{1}1} B^0_{\bar{2}00} B^0_{020}  \nonumber \\
&&
+ 6A^0_{\bar{1}11} A^0_{\bar{1}\bar{1}1} B^0_{200} B^0_{00\bar{2}}
+ 6A^0_{11\bar{1}} A^0_{1\bar{1}\bar{1}} B^0_{\bar{2}00} B^0_{002}  \nonumber \\
&&
+ 6A^0_{1\bar{1}1} A^0_{\bar{1}\bar{1}1} B^0_{00\bar{2}} B^0_{020}
+ 6A^0_{11\bar{1}} A^0_{\bar{1}1\bar{1}} B^0_{002} B^0_{0\bar{2}0}. \nonumber
\eea
In general, the above free-energy density is a multi-variate function of the fourteen amplitudes $A_i^0$ and $B_i^0$ of 
different density waves, and cannot be represented graphically in a simple way. However, the free-energy barrier between solid and liquid can be made explicit by assuming that all the $\langle 111 \rangle$ and all the $\langle 200 \rangle$ density waves have the same amplitude (i.e., $A_i^0=A$ and $B_i^0=B$, respectively).  In this isotropic approximation (see \cite{Wu07} for the bcc analog), the free-energy density
becomes
\bea
\label{eq:feneq}
&&\Delta f^{AE} =4 (-1+ 3{\psi_c}^2 ) A^2 
\nonumber \\
&&+3(-1+ 3{\psi_c}^2 +R(-Q_1^2+1)^2) B^2 \nonumber \\
&& + 54 A^4 +\frac{45}{2} B^4 + 144 A^2 B^2 + 72 {\psi_c} A^2B.
\eea
This expression can also be obtained by evaluating directly the difference between
the solid and liquid free-energy densities, 
$\Delta f^{AE}=\epsilon^{-2}(f_s-f_l)$, with $f_l$ and $f_s$ given by Eqs. (\ref{eq:fldef})
and (\ref{eq:fs}), respectively, and the substitutions 
$\bar\psi_s=\bar\psi_l=\psi_c\epsilon^{1/2}$, $A_s=A\epsilon^{1/2}$
and $B_s=B\epsilon^{1/2}$.
For the parameters $R=0$ and ${\psi_c}$ calculated from
the common tangent construction, we plot in Fig.~\ref{F_AB}
 the free-energy landscape 
as a function of
amplitudes $A$ and $B$.
The free-energy landscape exhibits two minima that correspond to the stable liquid
and solid phases. The above amplitude equation calculation shows that the two-mode PFC model
describes well solid-liquid coexistence with a well-defined 
free-energy barrier between solid and liquid. 

We have only treated here the case where the
amplitudes are spatially uniform to characterize the bulk free-energy landscape. A more general free-energy functional that
includes gradient terms would be necessary to treat the case where the amplitudes are spatially varying. Such a functional could then be used to compute the excess free-energy of the solid-liquid interface and its anisotropy, as done previously for bcc \cite{Wu07}. Those computations will be presented elsewhere.

\subsection{Relative stability of fcc and bcc}
\label{stabfccbcc}

\begin{figure}
\centering
\includegraphics[width=0.5\textwidth, angle=0]{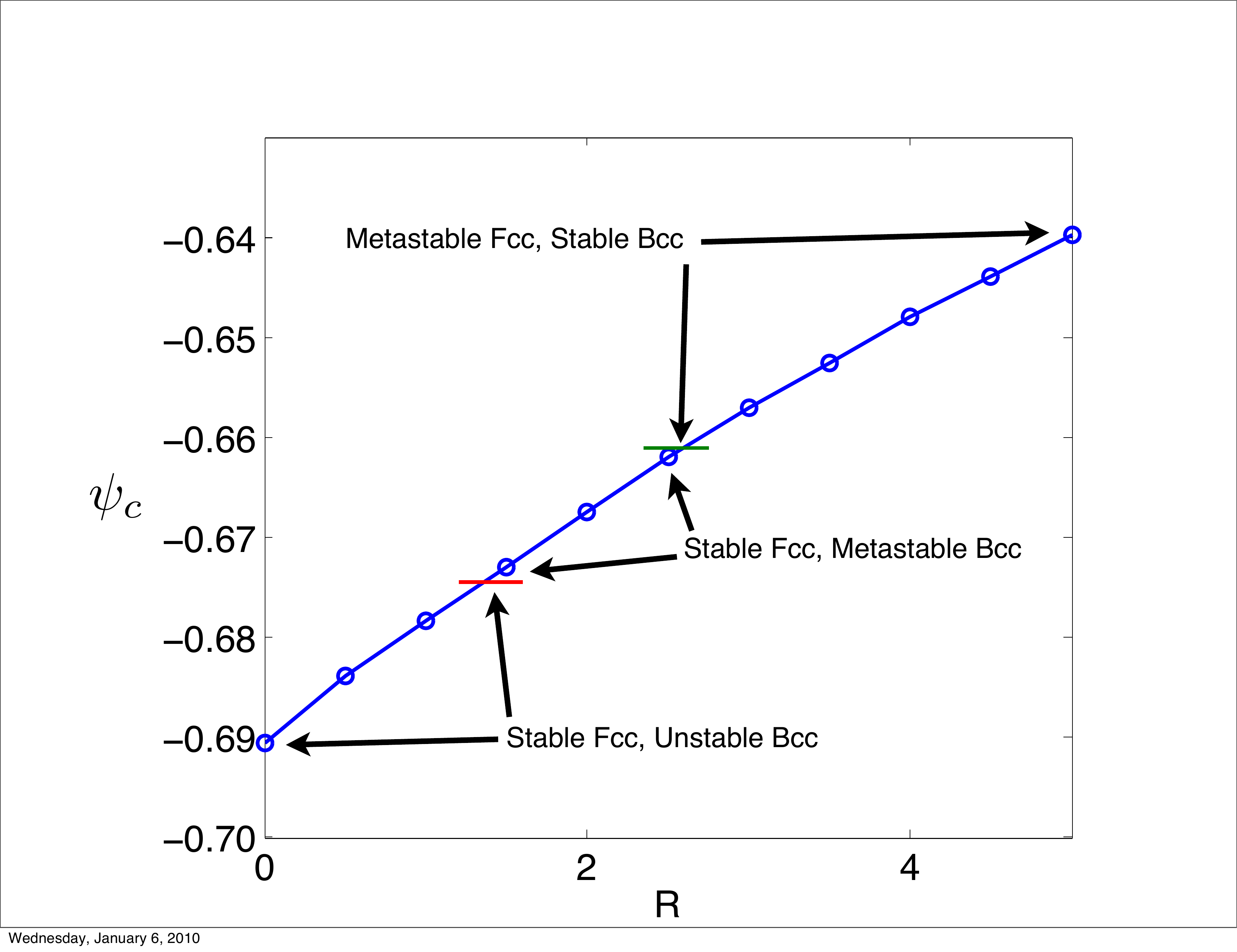}
\caption{(Color Online) $\psi_c$ as a function of $R$ showing different ranges of bcc or fcc metastability, stability, or instability.}
\label{rpsic}
\end{figure}

So far, we have only examined the possibility of fcc-liquid coexistence. However, the phase-diagram of Fig. \ref{phase} shows that bcc can have a lower free-energy than fcc for small enough $\epsilon$ if $R_1$ is finite. We now use the amplitude equations to understand the relative stability of fcc and bcc. As a first step, it is useful to re-examine the scaling of the mean density that is controlled by the parameter $\psi_c$. We computed previously the equilibrium solid and liquid densities using the common tangent construction, from which we obtained the scalings $(\bar \psi_s+\bar \psi_l)/2=\psi_c \epsilon^{1/2}$, which defines $\psi_c$, and $ \psi_s-\bar \psi_l\sim \epsilon^{3/2}$, which shows that the size of the density difference between solid and liquid can be neglected in the small $\epsilon$ limit. We can also compute $\psi_c$ 
more directly from Eq.~(\ref{eq:feneq}) by requiring  
\begin{equation}
\frac{\partial \Delta f^{AE}}{\partial A}=\frac{\partial \Delta f^{AE}}{\partial B}=0,\label{cond1}
\end{equation}
and
\begin{equation}
\Delta f^{AE}=0,\label{cond2}
\end{equation}
with all the above relations evaluated at the equilibrium values of $A$ and $B$ in the solid. Eq. (\ref{cond1}) stems from the requirement that the solid amplitudes must correspond to a free-energy minimum, which fixes those amplitudes uniquely as functions of $\psi_c$. Eq. (\ref{cond2}), in turn, is the requirement that the free-energies of solid and liquid must be equal in equilibrium, which fixes $\psi_c$ uniquely for a given $R$.  A plot of $\psi_c$ versus $R$ obtained in this way using Eqs. (\ref{cond1}) and (\ref{cond2}) is shown in Fig.~\ref{rpsic}.

\begin{figure}
\centering
\includegraphics[width=0.5\textwidth, angle=0]{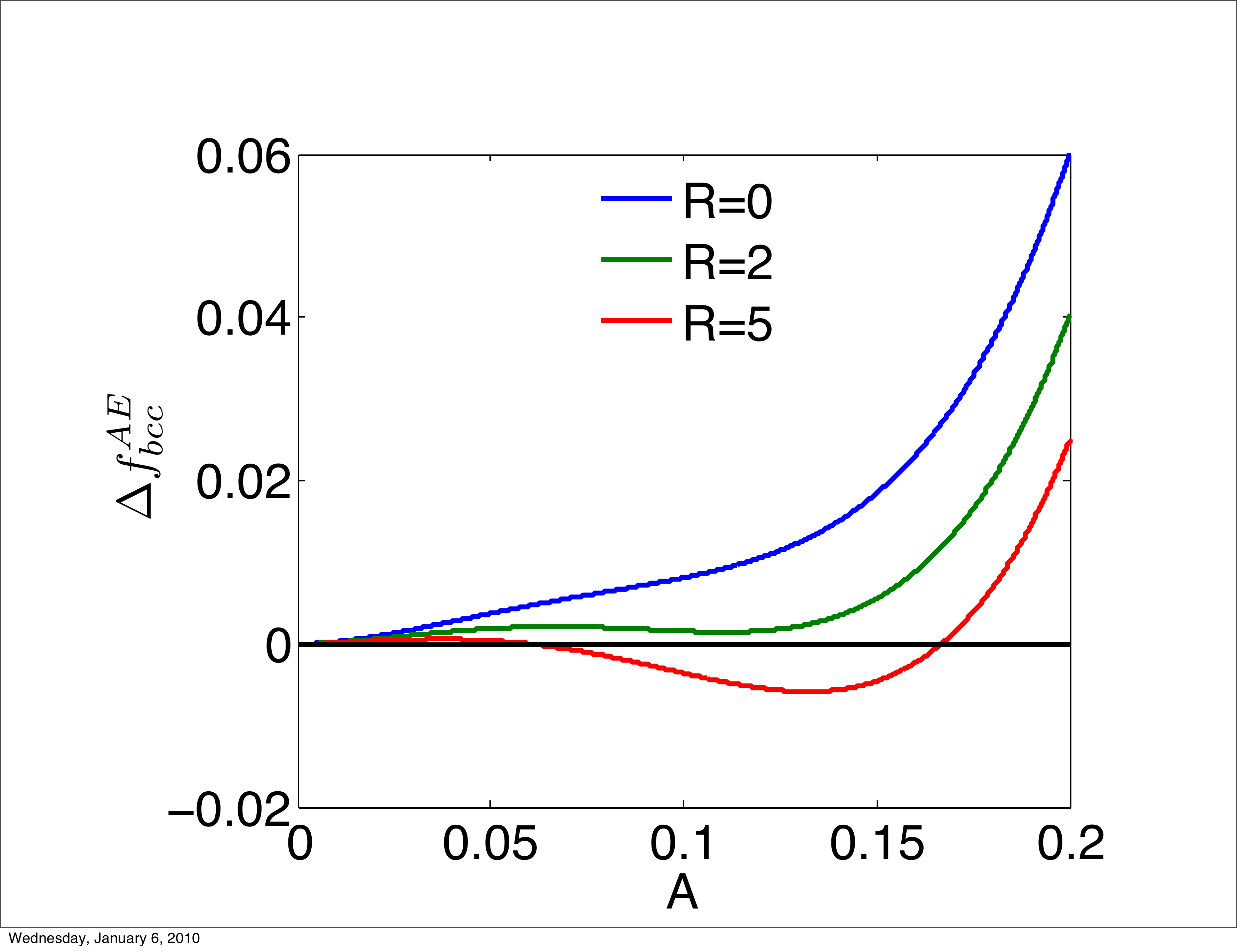}
\caption{Plots of free-energy density of bcc relative to the liquid, $\Delta f_{bcc}^{AE}$, as a function of the amplitude A of
$\langle 110 \rangle$ density waves for different values of $R$.  When the value of $\Delta f_{bcc}^{AE}$ corresponding to the solid bcc free-energy minimum is negative, bcc is favored over fcc.  As $R$ decreases, bcc first becomes metastable with respect to fcc and then unstable as the local solid free-energy minimum disappears.}
\label{fAEbcc}
\end{figure}

The relationship between $\psi_c$ and $R$, denoted by $\psi_c(R)$, can now be used to assess the relative stability of bcc and fcc.  To obtain an analogous expression to Eq. (\ref{eq:feneq}) for bcc, we substitute into the two-mode free-energy functional defined by Eqs. (\ref{Fdef}) and (\ref{twomode2}), the one-mode expansion of the bcc crystal density field in terms of principal set of $\langle 110 \rangle$ density waves
\bea
& &\psi(\vec{r})\approx \epsilon^{1/2}\psi_c +4\epsilon^{1/2} A(\mbox{cos}\,qx\,\mbox{cos}\,qy\nonumber\\
&&~~~~~~~~~~~~~+\mbox{cos}\,qx\,\mbox{cos}\,qz+\mbox{cos}\,qy\,\mbox{cos}\,qz),
\eea
where $q=1/\sqrt{2}$.
We obtain
\begin{equation}
\Delta f_{bcc}^{AE}=6(3\psi_c^2-1)A^2+48\psi_cA^3+135A^4,\label{fbcc}
\end{equation}
where $A$ now denotes the amplitude of $\langle 110 \rangle$ density waves and we have used the subscript ``bcc'' to distinguish this free-energy difference between bcc and liquid from the one between fcc and liquid, $\Delta f^{AE}$, defined by Eq. (\ref{eq:feneq}). By definition, $\Delta f^{AE}=0$ for solid fcc in equilibrium with the liquid. Therefore, to assess the relative stability of bcc and fcc, we can plot $\Delta f_{bcc}^{AE}$ defined by Eq. (\ref{fbcc}) versus $A$ and check if the value corresponding to the solid bcc minimum is above (below) zero in which case fcc (bcc) has a lower free-energy than bcc (fcc). Such plots shown in Fig.~\ref{fAEbcc} show that bcc becomes metastable with respect to fcc and then unstable (with the disappearance of the local solid bcc free-energy minimum) as $R$ is decreased. A detailed study as a function of $R$ shows that bcc first becomes metastable for $R<R_c$ where $R_c=2.68$ and then unstable as $R$ is decreased below a second threshold value ($\approx 1.43$), giving rise to the three different stability regimes as a function of $R$ shown in Fig. \ref{rpsic}. Translated in terms of the phase diagram constructed at fixed $R_1$, this implies that bcc becomes favored over fcc when $\epsilon<\epsilon_c$ where
\begin{equation}
\epsilon_c=\frac{R_1}{R_c}.
\end{equation}
For $R_1=0.05$, the above expression predicts $\epsilon_c\approx0.019$ that is in good quantitative agreement with the phase diagram of Fig.~\ref{phase}.  As $R_1$ increases, $\epsilon_c$ increases and the switch from stable bcc-liquid to fcc-liquid coexistence moves to higher values of $\epsilon$ in the phase diagram.

\section{Parameter determination}
 
In this section, we derive expressions to relate the two-mode PFC model parameters to material parameters by extending our previous approach for bcc \cite{Wu07}. As a first step, we match the peak liquid structure factor properties of the two-mode PFC model to the standard expression from classical DFT. The expression for the PFC liquid structure factor is obtained by
varying $\psi$ around its liquid value, $\psi=\bar \psi_l+\delta\psi$, and evaluating the corresponding variation $\Delta {\cal F}$ of the dimensional free-energy difference between solid and liquid using Eqs. (\ref{calF}) and (\ref{twomode}), and the
relation (\ref{eq:ampc}) between $\phi$ and $\psi$. Dropping terms of $\delta\psi$ higher than quadratic order, we obtain
\begin{eqnarray}
\Delta {\cal F}_{\rm PFC}&=&\frac{\lambda q_0^8}{g}\int d\rv \Bigg[\frac{\delta\psi}{2}[a+3\bar \psi_l^2\lambda q_0^8\nonumber \\
&+& \lambda (\nabla^2+q_0^2)^2((\nabla^2+q_1^2)^2+r_1)]\delta\psi\Bigg].
\end{eqnarray}
Substituting the Fourier transform,
\begin{equation}
\delta\psi=\int \frac{d\vec{k}}{(2\pi)^{3/2}}\delta\psi_ke^{i\vec{k}\cdot\vec{r}},\label{fourierpsi}
\end{equation}
we obtain
\begin{eqnarray}
\Delta {\cal F}_{\rm PFC}&=& \frac{\lambda q_0^8}{g}\int\int \frac{d\vec{k}d\vec{k}'}{(2\pi)^3}\frac{\delta\psi_k\delta\psi_{k'}}{2}
\Bigg\{[a+3\bar \psi_l^2\lambda q_0^8\nonumber \\
&+&\lambda(-k^2+q_0^2)^2((-k^2+q_1^2)^2+r_1)]\nonumber\\
&&\int d\vec{r}e^{i(\vec{k}+\vec{k}')\cdot\vec{r}}\Bigg\}\nonumber \\
&=& \frac{\lambda q_0^8}{g}\int d\vec{k}\frac{\delta\psi_k\delta\psi_{-k}}{2}
[a+3\bar \psi_l^2\lambda q_0^8\nonumber \\
&+&\lambda(-k^2+q_0^2)^2((-k^2+q_1^2)^2+r_1)].
\end{eqnarray}
A second expression for the
free-energy of a spatially inhomogeneous liquid is obtained from classic DFT
\bea
& &\Delta {\cal F}_{\rm DFT}=\frac{k_BT}{2}\int\int d\vec{r}d\vec{r}'\nonumber \\
& &\delta n(\vec{r})\left[ \frac{\delta(\vec{r}-\vec{r}')}{n_0}
-C(|\vec{r}-\vec{r}'|)\right]\delta n(\vec{r}'),\label{DFTeq}
\eea
where
\begin{equation}
\delta n(\vec{r})=n(\vec{r})-n_0 = \delta\phi(\vec{r})=\sqrt{\frac{\lambda q_0^8}{g}}\delta\psi(\vec r),\label{ndef}
\end{equation}
and
\begin{equation}
C(k)=n_0\int d\vec{r}C(|\vec{r}|)e^{-i\vec{k}\cdot\vec{r}},
\end{equation}
is the Fourier transform of the direct correlation function. Fourier transforming again, we obtain
\bea
\Delta {\cal F}_{\rm DFT}=\frac{\lambda q_0^8}{g}\frac{k_BT}{2n_0}\int d\vec{k} \delta\psi_k\delta\psi_{-k}\left[1-C(k)\right].
\eea
Equating $\Delta {\cal F}_{\rm PFC}=\Delta {\cal F}_{\rm DFT}$ and using the expression for
the liquid structure factor $S(k)=1/(1-C(k))$, we obtain
\begin{equation}
\label{eq:sk}
S(k)=\frac{k_BT}{n_0(a+3\lambda q_0^8\bar \psi_l^2+\lambda(-k^2+q_0^2)^2((-k^2+q_1^2)^2+r_1))}.
\end{equation}
By evaluating the above expression at the peak of the liquid structure factor, we obtain
\begin{equation}
a+3\lambda q_0^8\bar \psi_l^2=\frac{k_BT}{n_0S(q_0)},\label{aexp}
\end{equation}
or, using Eq.~(\ref{eq:epsc}) and the relationship $\bar \psi_l=\psi_c\epsilon^{1/2}$,  
\begin{equation}
\label{eq:epsilon1}
\epsilon = \frac{-k_BT}{n_0S(q_0)\lambda q_0^8(1-3\psi_c^2)}.
\end{equation}
A second relation is now needed to determine $\epsilon$ and $\lambda$ independently.
To obtain it, we substitute
Eq.~(\ref{eq:sk}) into the relation
$C(k)=(S(k)-1)/S(k)$
and compute the second derivative of $C(k)$ evaluated at
the peak of the liquid structure factor to obtain
\begin{equation}
\label{eq:epsilon2}
\lambda=-\frac{k_BTC''(q_0)}{8n_0q_0^6(\frac{1}{9}+\epsilon R)}.
\end{equation}
Eqs.~(\ref{eq:epsilon1}) and (\ref{eq:epsilon2}) combined now give  
\begin{equation}
\label{eq:epss}
\epsilon=\frac{8}{9(q_0^2S(q_0)C''(q_0)(1-3\psi_c^2)-8R)},
\end{equation}
and
\begin{equation}
\lambda=\frac{-9k_BTC''(q_0)}{8n_0q_0^6}+\frac{9k_BTR}{n_0S(q_0)(1-3\psi_c^2)q_0^8}.\label{lambda}
\end{equation}
In addition, the relation (\ref{ndef}) between the real and dimensionless densities expresses  
\begin{equation}
\label{eq:ggg}
g=\frac{\lambda q_0^8 {A_s^2}}{n_0^2u_s^2},
\end{equation}
in terms of the solid amplitude $A_s$ of the first $q_0$-mode. The two solid amplitudes $A_s$ and $B_s$ can be computed for a given $R$ by using the scaling relations $A_s=\epsilon^{1/2}A$ and $B_s=\epsilon^{1/2}B$ where $A$ and $B$ are 
the equilibrium values of the scaled amplitudes in solid. The latter are obtained, together with $\psi_c$ (Fig. \ref{rpsic}), by using the conditions (\ref{cond1}) and (\ref{cond2}) with $\Delta F^{AE}$ defined by Eq. (\ref{eq:feneq}).  

\begin{figure}
\centering 
\includegraphics[width=0.5\textwidth, angle=0]{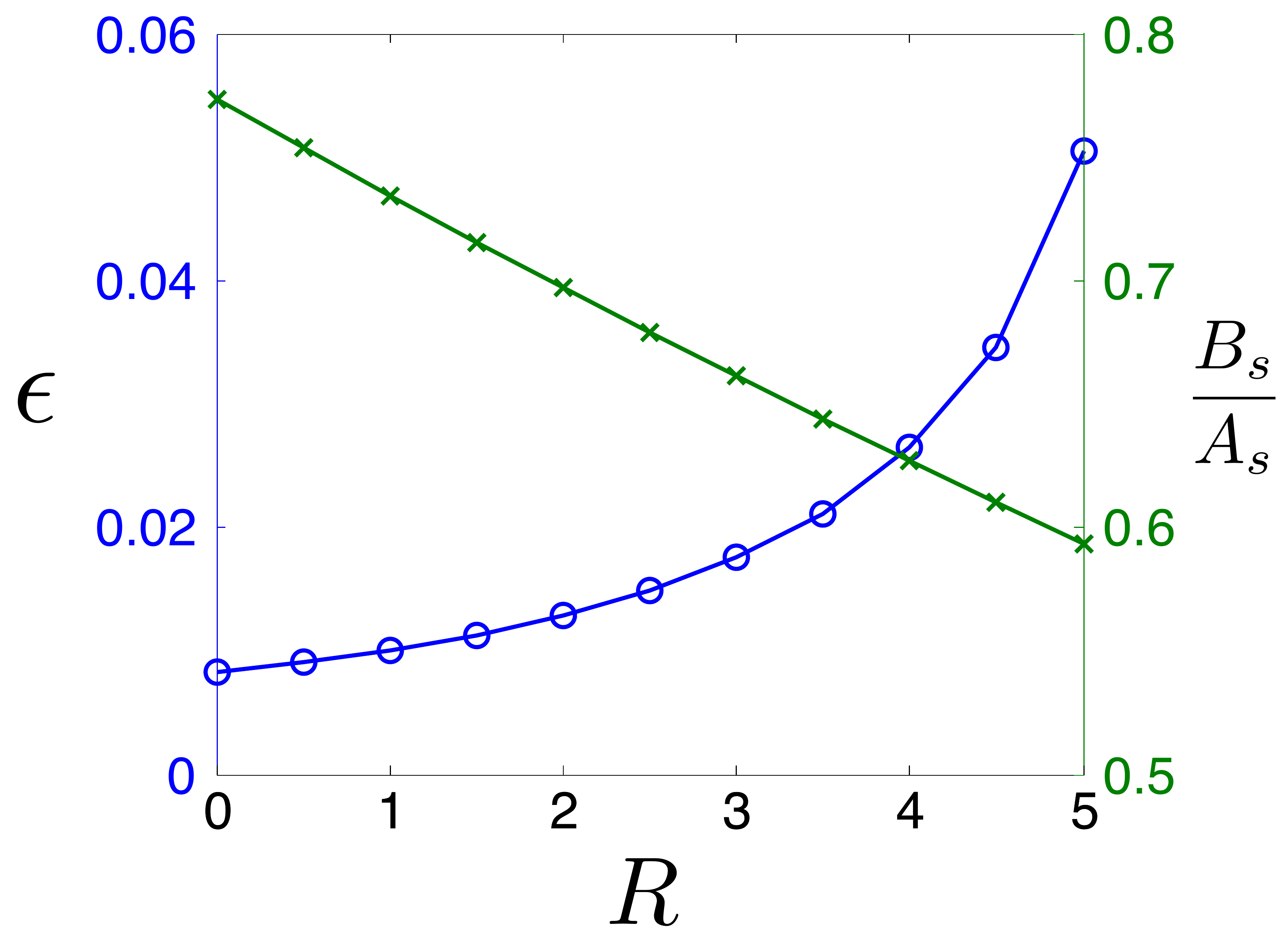}
\caption{(Color online) Plots showing the variation with $R$ of $\epsilon$ calculated by Eq.~(\ref{eq:epss}) and the ratio $B_s/A_s$ of the solid amplitudes of the second and first modes
calculated by the conditions (\ref{cond1}) and (\ref{cond2}) with $\Delta F^{AE}$ defined by Eq. (\ref{eq:feneq}).}
\label{epsabr}
\end{figure}

\begin{figure}
\centering 
\includegraphics[width=0.5\textwidth, angle=0]{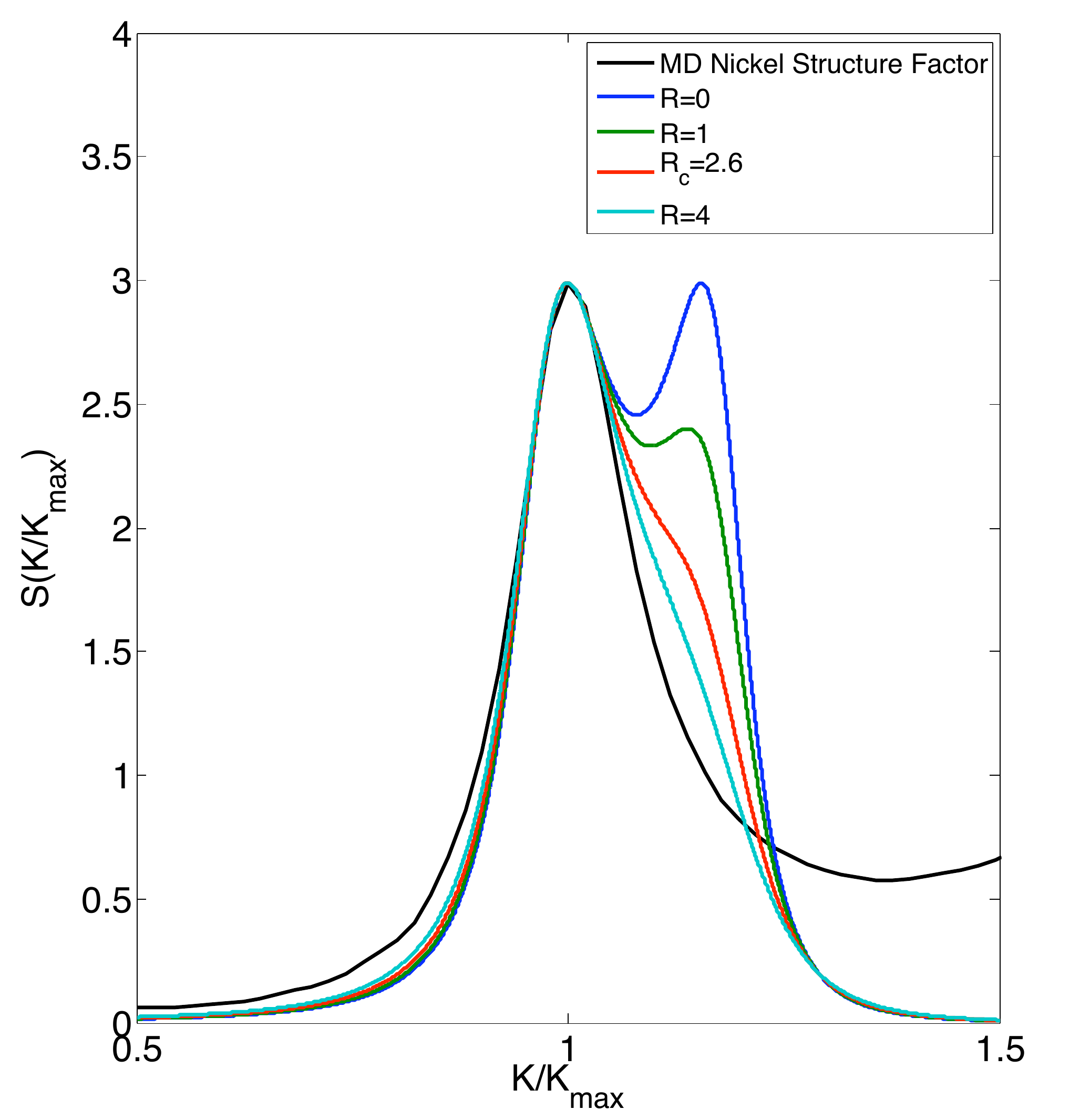}
\caption{(Color online) Liquid structure factor of the PFC model and from
MD simulations of pure Ni \protect\cite{Hoy06}.}
\label{sk}
\end{figure}

For a given $R$, Eqs. (\ref{eq:epss}), (\ref{lambda}), and (\ref{eq:ggg}) fix the three parameters $\epsilon$, $\lambda$, and $g$ of the PFC model uniquely in terms of peak liquid structure factor properties, $S(q_0)$ and $C''(q_0)$, where $q_0=|\vec K_{111}|$ here, and the solid density wave amplitude $u_s=u_{111}$. This still leaves the freedom to vary $R$ within the range where fcc is stable with respect to bcc (Fig. \ref{rpsic}). Varying $R$ changes the shape of the liquid structure factor as shown in Fig. \ref{sk} and decreasing $R$ below some threshold produces a second peak at $q_1=|\vec K_{200}|$, and generally increases the contribution of the second mode. Thus decreasing $R$ increases the amplitude of the second mode $u_{200}$ as shown in Fig. \ref{epsabr}. For simplicity, we used the value $R=0$ that yields a reasonable fit of this amplitude for pure Ni. The other input parameters computed by Hoyt \cite{Hoy06} using the EAM potential of Foiles, Baskes and Daw \cite{FBD} (FBD) are given in Table \ref{shift::table1}. The density wave amplitudes are calculated using the relation $u_i=\exp(-K_i^2/4a)$, which assumes that the crystal density field is a sum of Gaussians centered around each fcc lattice site. The value of $a$ is obtained from the expressions for the root-mean-square displacement of atoms in the solid $\sqrt{<|\vec{r}|^2>}=3/(2a)$ derived from this density field. For the value $\sqrt{<|\vec{r}|^2>}\approx 0.298 \AA$ from MD simulations, we obtain $u_{111}=\exp(-K_{111}^2/4a)=0.6639$
and $u_{200}=\exp(-K_{111}^2/4a)=0.5791$.

It should be noted that, with the present fitting procedure, the two-mode PFC model only reproduces the correct shape of the main peak of the liquid structure factor. The second peak is spurious and is only used to increase the amplitude of the second mode to some desired value. Since the second mode is critical to obtain solid-liquid coexistence, the lack of realism of the structure factor outside of the first peak is a limitation of the present two-mode model. The liquid structure factor could in principle be made more realistic by shifting the second peak to larger wavector and reducing its amplitude, which would couple the principal $\langle 111 \rangle$ RLVs to other sets such as $\langle 222 \rangle$ and $\langle 311 \rangle$. However, larger $k$-modes require a finer mesh and are computationally more costly to resolve. Whether such a fit would offer specific advantages remains to be investigated.

\begin{table}
\begin{center}
\caption{Input parameters for the PFC model computed from MD simulations of pure Ni \cite{Hoy06} using the FBD EAM potential \cite{FBD} and
corresponding PFC parameters.}
\begin{tabular}{c|c|c|c}
\hline
\hline
MD input parameters & Value & PFC parameters & Value \\
\hline
$q_0=K_{111}$ ($\AA^{-1}$) & $3.0376$ & $\lambda$ (eV$\AA^{11}$)& $0.026$  \\
$n_0$ ($\AA^{-3}$) & $0.0801$ &$g$ (eV$\AA^9$)& $8.53$ \\
$C''(q_0)$ ($\AA^{2}$)& $-9.1579$ &$\psi_c$ & $-0.6901$ \\
$S(q_0)$ & $2.9898$ &$R$ & $0$\\
$T_M$ (K)& $1811$ &$\epsilon$ & $0.00823$   \\
$u_{111}$ & $0.6639$ &$u_{111}$ & $0.6639$   \\
$u_{200}$ & $0.5791$ &$u_{200}$ & $0.5136$  \\
\hline
\hline
\end{tabular}
\label{shift::table1}
\end{center}
\end{table}

\section{Elastic constants}
\label{elasticity}

In this section, we derive analytical expressions for the elastic constants of the two-mode PFC model. We compare the results to MD computations of elastic constants at the melting point for parameters of fcc Ni. For completeness, we also carry out the same comparison for the standard one-mode PFC model for parameters of bcc Fe. Following the same approach as in Ref. \cite{Eldetal04}, we obtain the elastic constants by deforming the lattice from its ideal structure and computing the corresponding change of free-energy density.  We consider three different deformations 
\begin{eqnarray}
\label{eq:p1}
\psi_1(\vec{r})&=&\psi(x/(1+\xi),y/(1+\xi),z/(1+\xi)),\\
\psi_2(\vec{r})&=&\psi(x/(1+\xi),y/(1-\xi),z),\\
\psi_3(\vec{r})&=&\psi(x+\xi y,y,z),
\end{eqnarray} 
of the two-mode crystal density field  
We compute the change of free-energy density
\begin{equation}
\Delta f_i=\frac{F_i}{V_i}-f_s,\label{eq:p2}
\end{equation}
where $f_s$ is the free-energy density of the unperturbed solid 
and $F_i$ is the free-energy integrated over the
perturbed unit cell of volume $V_i$ with
\begin{eqnarray*}
\frac{F_1}{V_1}&=&\frac{1}{V(1+\xi)^3}\int_0^{a(1+\xi)} \int_0^{a(1+\xi)}\int_0^{a(1+\xi)}f(\psi_1(\vec{r}))dV,\\
\frac{F_2}{V_2}&=&\frac{1}{V(1-\xi
^2)}\int_0^{a} \int_0^{a(1-\xi)}\int_0^{a(1+\xi)}f(\psi_2(\vec{r}))dV,\\
\frac{F_3}{V_3}&=&\frac{1}{V}\int_0^{a} \int_0^{a}\int_{-\xi y}^{a-\xi y}f(\psi_3(\vec{r}))dV,
\end{eqnarray*}
where $dV=dxdydz$, $a$ is the lattice spacing, and $V=a^3$ is the unperturbed unit cell volume.

\subsection{fcc elastic constants for the two-mode model}

Using Eqs. (\ref{eq:p1})-(\ref{eq:p2}) with the two-mode crystal density field 
$\psi(\vec{r})$ defined by Eq. (\ref{eq:fccs})
and the free-energy density $f(\psi(\vec{r}))$ defined by Eq. (\ref{twomode2}), we obtain
the dimensionless elastic constants
\begin{eqnarray}
\label{eq:f1}
\Delta f_1=\left(\frac{3}{2}\tilde{C}_{11}+3\tilde{C}_{12}\right)\xi^2=(\alpha+\beta)\xi^2,\\
\Delta f_2=(\tilde{C}_{11}-\tilde{C}_{12})\xi^2=\frac{2}{3}\beta\xi^2,\\
\Delta f_3=\frac{\tilde{C}_{44}}{2}\xi^2=(\alpha/9+\delta)\xi^2,
\end{eqnarray}
where we have defined
\begin{eqnarray}
\alpha&=&16(1-2Q_1^2+Q_1^4+R_1)A_s^2,\\
\beta&=&\frac{8}{27}(284-315Q_1^2+81Q_1^4+81R_1)B_s^2,\\
\delta&=&\frac{8}{9}R_1B_s^2,
\end{eqnarray}
We can set $R_1\approx 0$ in the above expression since $R_1\ll 1$ for typical model parameters where the fcc lattice is favored. With only the principle
$\langle 111\rangle$ RLVs ($B_s=0$), all three elastic constants are equal
$\tilde{C}_{11}=\tilde{C}_{12}=\tilde{C}_{44}=2\alpha/9$, which gives a vanishing tetragonal shear modulus
$\tilde{C}'=(\tilde{C}_{11}-\tilde{C}_{22})/2$. The inclusion of the $\langle 200\rangle$ RLVs, however, raises the value of $\tilde{C}_{11}$, which becomes
%\[\tilde{C}_{11}=\frac{2}{9}\alpha\left(1+\frac{1}{2}\frac{(8+81R_1)}{(1+9R_1)}\frac{B_s^2}{A_s^2}\right),\]
\[\tilde{C}_{11}=\frac{2}{9}\alpha\left(1+4\frac{B_s^2}{A_s^2}\right),\]
while leaving the values of $\tilde{C}_{12}$ and $\tilde{C}_{44}$ unchanged, thereby making $\tilde{C}'$ finite as desired. 
 Finally, converting back to dimensional units using the relation 
$C_{ij}=(\lambda^2q_0^{16}/g)\tilde{C}_{ij}$, we obtain
\bea
C_{11}=-\frac{4}{9}n_0k_BTC''(K_{111})q_0^2u_{111}^2\left(1+4\frac{u_{200}^2}{u_{111}^2}\right),
\eea
and
\bea
C_{12}=C_{44}=-\frac{4}{9}n_0k_BTC''(K_{111})q_0^2u_{111}^2.
\eea
The elastic constants computed with the parameters of Table~\ref{shift::table1} are compared to the predictions of MD simulations in Table \ref{shift::table2}. The MD simulations for fcc Fe and bcc Ni were carried out using the EAM potentials from Mendelev, Han, Srolovitz, Ackland, Sun
and Asta MH(SA)$^2$ \cite{MHSA2}, and Foiles, Baskes and Daw \cite{FBD}, respectively. The same EAM potentials
were used to compute the input parameters for the PFC model and the elastic constants. The input parameters for Fe are the
same as in Ref. \cite{Wuetal06}. The input parameters for Ni were computed by Hoyt \cite{Hoy06}. The elastic constants for both
Fe and Ni were computed by Foiles \cite{Foi09}. Their values at the melting point are smaller than at zero
temperature as shown by Foiles for a different Ni EAM potential \cite{Foi10}.

\begin{table}
\begin{center}
\caption{Comparison of elastic constants at the melting point predicted by the two-mode PFC model for fcc Ni and the one-mode PFC model for bcc Fe and MD simulations \protect\cite{Foi09}.}
\begin{tabular}{c|c|c|c|c}
\hline
\hline
Quantity & PFC fcc & MD fcc & PFC bcc & MD bcc  \\
\hline
$C_{11}$ (GPa) & 106.6 & 155.4  & 90.0 & 128.0 \\
$C_{12}$ (GPa) & 31.4 & 124.7 & 45.0 & 103.4 \\
$C_{44}$ (GPa) & 31.4 & 66.0 & 45.0 & 63.9  \\
Bulk Modulus (GPa) & \multirow{2}{*}{56.5} & \multirow{2}{*}{134.9} &  \multirow{2}{*}{60.0} &   \multirow{2}{*}{111.6} \\
$(C_{11}+2C_{12})/3$&  & & & \\
\hline
\hline
\end{tabular}
\label{shift::table2}
\end{center}
\end{table}

\subsection{bcc elastic constants for the one-mode model}

For the one-mode model, we use again
Eqs. (\ref{eq:p1})-(\ref{eq:p2}) with the one-mode bcc crystal density field 
\begin{equation}
\label{eq:onemode}
\psi(\vec{r})\approx\bar{\psi} +4A_s(\mbox{cos}\,qx\,\mbox{cos}\,qy+\mbox{cos}\,qx\,\mbox{cos}\,qz+\mbox{cos}\,qy\,\mbox{cos}\,qz).
\end{equation} 
where $q=1/\sqrt{2}$ and the free-energy density $f(\psi(\vec{r}))$ defined by Eq. (\ref{onemode2}). We obtain
\begin{eqnarray}
\Delta f_1&=&\left(\frac{3}{2}\tilde{C}_{11}+3\tilde{C}_{12}\right)\xi^2=\alpha_{bcc} \xi^2,\\
\Delta f_2&=&(\tilde{C}_{11}-\tilde{C}_{12})\xi^2=\frac{\alpha_{bcc}}{6}\xi^2,\\
\Delta f_3&=&\frac{\tilde{C}_{44}}{2}\xi^2=\frac{\alpha_{bcc}}{12}\xi^2,
\end{eqnarray}
where $\alpha_{bcc}=24A_s^2$.
This yields the dimensionless elastic constants
$
\tilde{C}_{11}=2\,\tilde{C}_{12}=2\,\tilde{C}_{44}=8A_s^2.
$
Finally, converting back to dimensional units using the relation 
$C_{ij}=(\lambda^2q_0^8/g)\tilde{C}_{ij}$, we obtain
\begin{equation}
C_{11}=2\,C_{12}=2\,C_{44}=-n_0k_BTC''(K_{110})q_{0}^2u_{110}^2.
\end{equation}
The elastic constants computed with the input parameters of Table~\ref{shift::tablebcc} for bcc Fe are compared to the predictions of MD simulations in Table \ref{shift::table2}.

\begin{table}
\begin{center}
\caption{Input parameters for the PFC model computed from MD simulations of pure Fe \cite{Wuetal06} using the EAM potential from MH(SA)$^2$ \cite{MHSA2} and
corresponding PFC parameters.}
\begin{tabular}{c|c|c|c}
\hline
\hline
MD input parameters & Value & PFC parameters & Value \\
\hline
$q_0=K_{110}$ ($\AA^{-1}$) & $2.985$ & $\lambda$ (eV$\AA^{7}$)& $0.291$  \\
$n_0$ ($\AA^{-3}$) & $0.0765$ &$g$ (eV$\AA^9$)& $9.705$ \\
$C''(q_0)$ ($\AA^{2}$)& $-10.40$ &$\epsilon$&  $0.0923$ \\
$S(q_0)$ & $3.012$ & $u_{110}$& $0.72$ \\
$T_M$ (K)& $1771$ & &   \\
$u_{110}$ & $0.72$ & &   \\
\hline
\hline
\end{tabular}
\label{shift::tablebcc}
\end{center}
\end{table}

\section{Two-dimensional square lattice}

\begin{figure}
\centering 
\includegraphics[width=0.4\textwidth, angle=0]{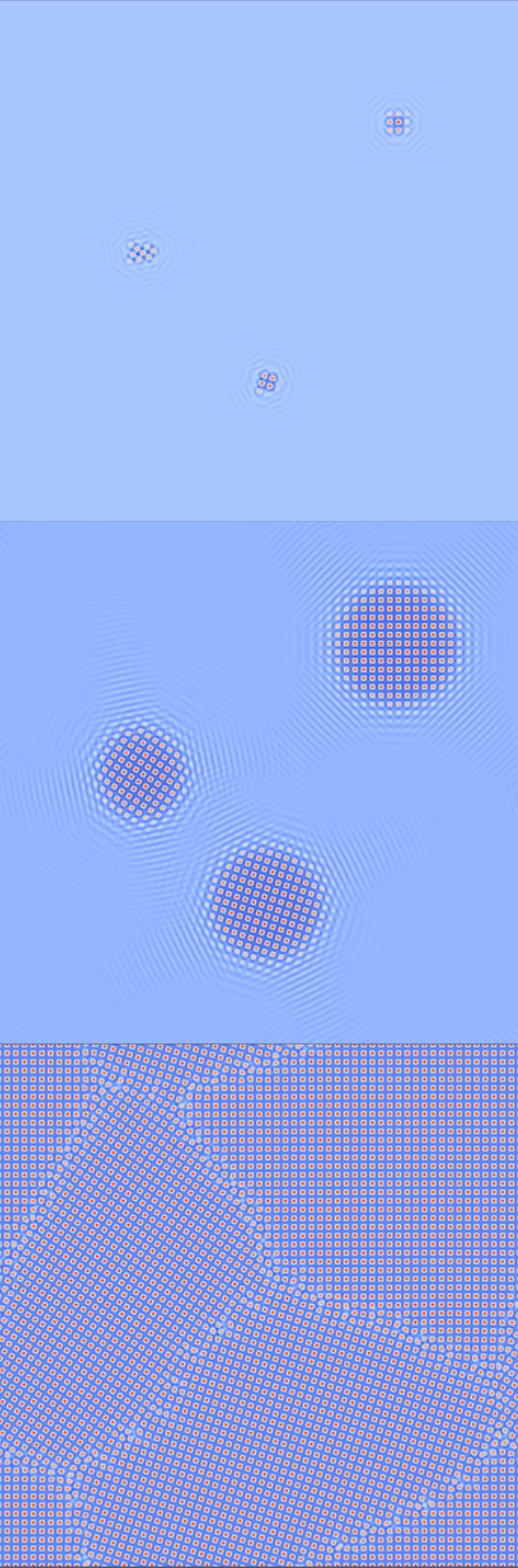}
\caption{(Color online) Example of polycrystalline solidification for two-dimensional square lattices. The snapshots are at dimensionless times $t=10,100,\mbox{and}\,1000$.  The parameters are $\epsilon=0.15$, $R_1=0$, $Q_1=\sqrt{2}$, and $\bar{\psi}=-0.23$.}
\label{sqpic}
\end{figure}

As an example of application of the two-mode model to other lattice structures, we briefly examine the example of two-dimensional square lattices. Those lattices are obtained by coupling $\langle 10 \rangle$ and $\langle 11 \rangle$ density waves with $Q_1=\sqrt{2}$, as demonstrated previously by Lifshitz and Petrich \cite{LifPet97} for a modifed Swift-Hohenberg model that corresponds to the $R_1=0$ limit of the present two-mode model. The liquid free-energy density is given by
\begin{equation}
f^{sq}_l(\bar{\psi}_l)=-(\epsilon-4-R_1)\frac{\bar\psi_l^2}{2}+\frac{\bar\psi_l^4}{4},
\label{eq:flsq}
\end{equation}
and the solid free-energy density is obtained by substituting the two-mode crystal density field
\begin{equation}
\psi(\vec{r})\approx \bar\psi + 2A_s(\cos{x}+\cos{y}) + 4B_s (\cos{x}\cos{y})
\end{equation}
into the free-energy functional defined by Eqs. (\ref{Fdef}) and (\ref{twomode2}) with $Q_1=\sqrt{2}$ , which yields
\begin{eqnarray}
f^{sq}_s(\bar\psi_s) &=& 2(-\epsilon + 3{\bar{\psi}_s}^2 ) {A_s}^2 
+2(-\epsilon +  3 {\bar{\psi}_s}^2 + R_1)  {B_s}^2 \nonumber \\ 
&&+24 \bar{\psi}_s {A_s}^2 B_s 
+36{A_s}^2 {B_s}^2   +9{A_s}^4 + 9{B_s}^4 \nonumber \\
&&-\frac{\epsilon}{2} {\bar{\psi}_s}^2 + \frac{R_1}{2} {\bar{\psi}_s^2}
+2{\bar{\psi}_s}^2 + \frac{1}{4} {\bar{\psi}_s}^4.
\label{eq:fssq}
\end{eqnarray}
For $R_1=0$, we obtain $\psi_c=-0.6782$ numerically from a log-log plot of the mean equilibrium density versus $\epsilon$ similar to Fig. \ref{fcc_liq_avg_psi} which is determined from the common tangent construction. The feasibility of the two-model to model polycrystalline solidification and grain boundaries is illustrated in Fig. \ref{sqpic}. As for fcc, the second mode turns out to be essential to obtain physically meaningful elastic constants. Following the same procedure as for fcc in the last section (with deformations of the unit cell now constrained to the $x-y$ plane) we obtain
\begin{eqnarray}
\Delta f_1^{sq}&=&\left(\tilde{C}_{11}+\tilde{C}_{12}\right)\xi^2=(\alpha_{sq}+\beta_{sq})\xi^2,\\
\Delta f_2^{sq}&=&(\tilde{C}_{11}-\tilde{C}_{12})\xi^2=(\alpha_{sq}+\Lambda_{sq})\xi^2,\\
\Delta f_3^{sq}&=&\frac{\tilde{C}_{44}}{2}\xi^2=\delta_{sq}\xi^2,
\end{eqnarray}
with
\begin{eqnarray}
\alpha_{sq}&=&8(1-2Q_1^2+Q_1^4+R_1)A_s^2,\\
\beta_{sq}&=&8(58-37Q_1^2+5Q_1^4+5R_1)B_s^2,\\
\Lambda_{sq}&=&8R_1B_s^2,\\
\delta_{sq}&=&4(32-21Q_1^2+3Q_1^4+3R_1)B_s^2.
\end{eqnarray}
Again we look in the limit  $R_1\approx 0$. This yields the dimensionless elastic constants
\begin{eqnarray}
\tilde{C}_{11}&=&\alpha_{sq}+\beta_{sq}/2,\\
\tilde{C}_{12}&=&\beta_{sq}/2,\\
\tilde{C}_{44}&=&2\delta.
\label{eq:sqelastic2}
\end{eqnarray}
These relations show that the one-mode crystal density field consisting only of a superposition of $\langle 10 \rangle$ density waves ($B_s=0$) yield vanishing shear moduli, which become finite with the inclusion of the second mode.

\section{Concluding remarks}

In summary, we have presented a two-mode PFC model with a phase-diagram that includes
different temperature ranges for bcc-liquid and fcc-liquid coexistence. The relative sizes of these ranges can be
changed by varying one model parameter that controls the relative magnitudes of the amplitudes of the
two modes, corresponding to [111] and [200] density waves, respectively. 
We have shown that the free-energy
landscape for fcc-liquid coexistence has a double-well structure with a finite free-energy barrier between solid and liquid  
in the plane of the amplitudes of the two modes. We have demonstrated the feasibility of the model with some numerical
examples of fcc polycrystalline growth and twin growth, as well as for two-dimensional square lattices. 

At a more quantitative level, we have determined the model parameters by fitting the peak liquid structure
factor properties ($S(q_0)$ and  $C''(q_0)$) and solid-density wave amplitudes as an extension
of our previous study of bcc Fe \cite{Wu07}. Furthermore,
we have derived analytical expressions for the elastic constants. With input values for those parameters
from MD simulations of pure Ni, we have found that the PFC model elastic constants
are in reasonable agreement with MD results given the simplicity of the model, which
neglects the contributions of many other modes that are present in a realistic description of the
crystal density field. Those expressions also stress the necessity of having at least
two distinct modes to obtain physically meaningful values of the elastic constants for fcc in the physically relevant
small $\epsilon$-limit of the PFC model, which
is also true for square lattices.
We have found that the standard one-mode PFC model also predicts reasonable
values of the elastic constants for pure bcc Fe, and we have argued that any one- or two-mode model
will predict similar elastic constants for bcc and fcc with the same peak liquid structure factor properties 
and solid density wave amplitudes

Finally, while the numerical examples focused on crystal growth, it might also be possible to use the two-mode PFC model to study the bcc/fcc martensitic transformation, which has been modeled by other phase-field approaches that make use of structural order parameters (see Refs.  \cite{Wang09,Wang06} and references therein). The ability to vary the relative stability of fcc and bcc crystal structures, which was demonstrated here, should prove particularly useful for this application.

\vspace{1cm}

\begin{acknowledgments} 
This work was supported by DOE grant DE-FG02-07ER46400 and the DOE sponsored Computational Materials Science Network program. We thank Mark Asta and Jeff Hoyt for valuable exchanges and Stephen Foiles for providing values of the elastic constants computed from molecular dynamics simulations. \end{acknowledgments} 

\appendix
\section{Twin boundary energy}

\begin{figure}
\vspace{0.5cm}
\centering 
\includegraphics[width=0.5\textwidth, angle=0]{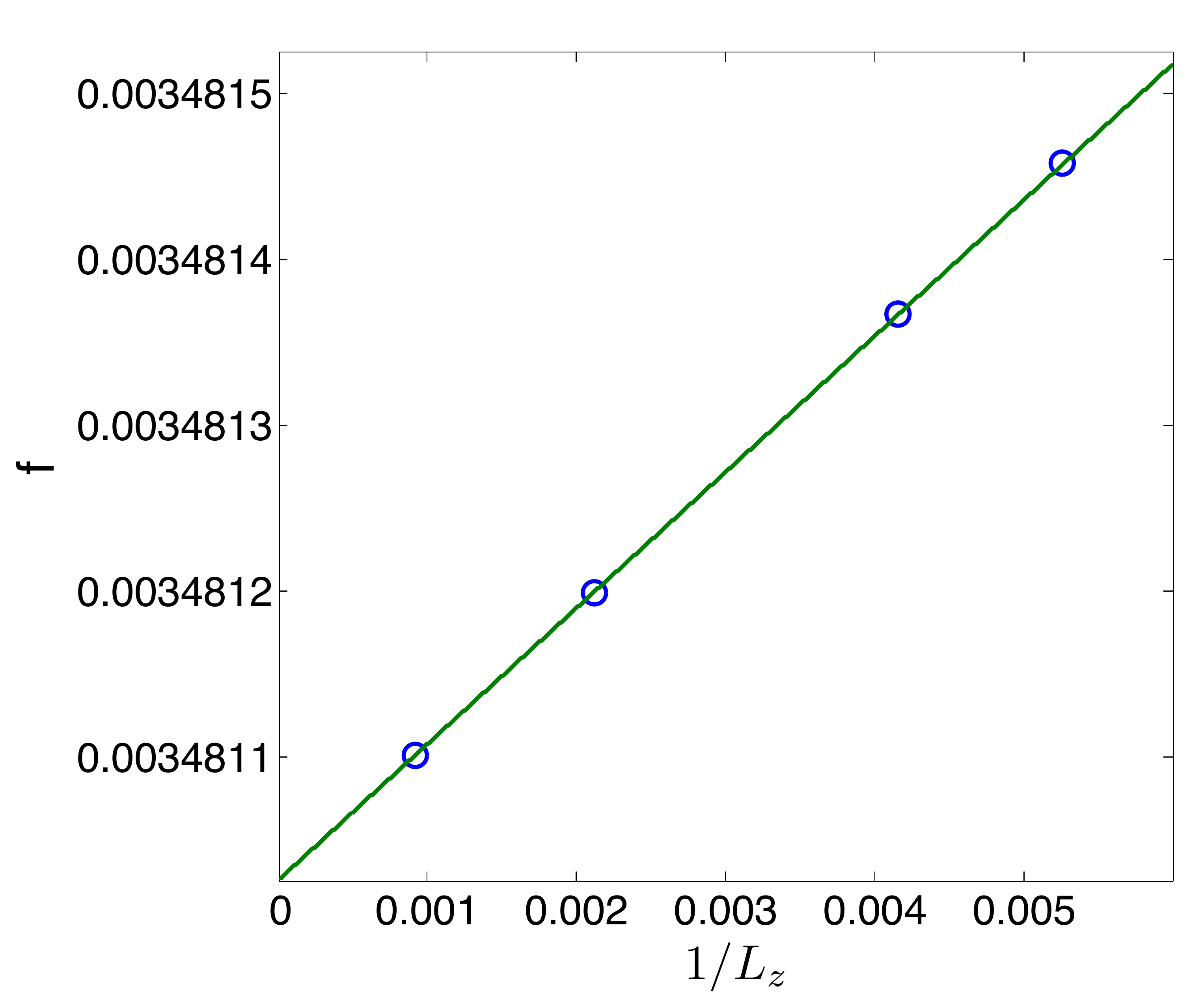}
\caption{(Color online) Plot of free-energy density versus inverse of the system length perpendicular to the twin boundary used to compute its excess free-energy.}
\label{inv}
\end{figure}

We computed the coherent (111) twin boundary energy using the method put forth in Ref. \cite{Mathis}, which exploits the dependence of the free-energy on system size. We performed simulations for four different lengths $L_z$ along the axis perpendicular to the boundary. By plotting the bulk free energy density $f$ against the inverse of this length (Fig.~{\ref{inv}}), we then extracted the boundary energy from the slope of this plot using the relation 
\begin{equation}
f=f_s(\bar{\psi})+2\frac{\tilde{\gamma}_{twin}}{L_z},
\end{equation}
where $f_s(\bar{\psi})$ is the free energy density of a perfect crystal.  This method turns out to be more accurate than computing directly the excess free-energy of the boundary for a fixed system size \cite{Mathis}. This calculation gives twice the boundary energy since there are two boundaries in our periodic system.  We convert the result to dimensional units through, $\gamma_{twin}=(\lambda^2 q_0^{15}/g) \tilde{\gamma}_{twin}$, which yields the reasonable value of 29.9 mJ/m$^2$ for the same Ni parameters used to compute the elastic constants.

\end{document}